\documentclass[12pt,USenglish]{article}
\pdfoutput=1

\usepackage{setspace}
\usepackage{simplewick}
\usepackage[dvipsnames]{xcolor}
\usepackage[normalem]{ulem}
\usepackage{cite}
\usepackage{dsfont}
\usepackage{marginnote}
\usepackage{placeins}

\renewcommand{\Im}{{\rm Im \, }}

\newcommand{\str}{\mathrm{STr}}
\newcommand{\tr}{\mathrm{Tr}}
\newcommand{\co}{\mathrm{c}}
\newcommand{\fl}{\mathrm{f}}

\newcommand{\xdownarrow}[1]{%
  {\left\downarrow\vbox to #1{}\right.\kern-\nulldelimiterspace}
}

\usepackage[utf8]{inputenc}
\usepackage{geometry}
\usepackage[USenglish]{babel}
\usepackage{verbatim}
\usepackage{amsmath}
\usepackage[normalem]{ulem}
\usepackage{amssymb}
\usepackage{graphicx}
\usepackage{setspace}
\usepackage{esint}
\usepackage{verbatim}

\usepackage{color}
\definecolor{darkgreen}{rgb}{0,0.5,0}
\definecolor{darkblue}{rgb}{0,0,0.6}
\definecolor{purple}{rgb}{0.4,.2,0.7}
\definecolor{orange}{rgb}{0.95, 0.5, 0.3}

\usepackage[colorlinks=true,citecolor=darkgreen,linkcolor=darkblue,urlcolor=blue]{hyperref}

\numberwithin{equation}{section}
\numberwithin{table}{section}

\def\cV{{\cal V}}
\def\cH{{\cal H}}

\def\be{\begin{equation}}
\def\ee{\end{equation}}
\def\bea{\begin{eqnarray}}
\def\eea{\end{eqnarray}}
\def\ba{\begin{align}}
\def\ea{\end{align}}

\def\id{1_{2\times 2}}

\makeatletter
\newcommand{\vast}{\bBigg@{4}}
\newcommand{\Vast}{\bBigg@{5}}
\makeatother

\begin{document}
\begin{spacing}{1.1}
  \setlength{\fboxsep}{3.5\fboxsep}

~
\vskip5mm

\begin{center} {\Large \bf Late time physics of holographic quantum chaos}

\vskip10mm

Alexander Altland$^{1}$ \& Julian Sonner$^{2}$\\
\vskip1em
{\small
{\it 1) Institut für theoretische Physik, Zülpicher Str. 77, 50937 Köln, Germany} \\
\vskip2mm
{\it 2) Department of Theoretical Physics, University of Geneva, 24 quai Ernest-Ansermet, 1211 Gen\`eve 4, Suisse}
}
\vskip5mm

\tt{ alexal@thp.uni-koeln.de, julian.sonner@unige.ch}

\end{center}

\vskip10mm

\begin{abstract}

Quantum chaotic systems are often defined via the assertion that their spectral
statistics coincides with, or is well approximated by, random matrix theory. In this
paper we explain how the universal content of random matrix theory emerges as the
consequence of a simple symmetry-breaking principle and its associated Goldstone
modes. This allows us to write down an effective-field theory (EFT) description of
quantum chaotic systems, which is able to control the level statistics up to an
accuracy ${\cal O} \left(e^{-S} \right)$ with $S$ the entropy. We explain how the EFT
description emerges from explicit ensembles, using the example of a matrix model with
arbitrary invariant potential, but also when and how it applies to individual quantum
systems, without reference to an ensemble. Within AdS/CFT this gives a general
framework to express correlations between ``different universes'' and we explicitly
demonstrate the bulk realization of the EFT in minimal string theory where the
Goldstone modes are bound states of strings stretching between bulk spectral branes.
We discuss the construction of the EFT of quantum chaos also in higher dimensional
field theories, as applicable for example for higher-dimensional AdS/CFT dual pairs.

\end{abstract}

\pagebreak
\pagestyle{plain}

\setcounter{tocdepth}{2}
{}
\vfill
\tableofcontents
\section{Introduction and summary}
Important progress in lower-dimensional models of AdS/CFT duality has given renewed impetus to contemplate ensemble averages of boundary theories, which either entirely capture the bulk quantum gravity path integral, or at least agree with the latter in an averaged sense.

Ensemble averages are also known to play a crucial role in the understanding of chaotic quantum systems. Generic chaotic quantum systems are believed to exhibit behavior at late times that is well captured by that of a suitable random-matrix ensemble. Suitable here means that there is a small number of universality classes distinguished by the presence or absence of anti-unitary symmetries in the original quantum system and `late times' means `later than the Thouless time', whose precise definition we give below. Equivalently, this RMT universality finds its expression in the distribution of energy levels, in the sense that the energy levels of a chaotic quantum system exhibit the same statistical properties as those of the corresponding ensemble. 

A key question in both contexts is the relationship between the {\it individual, fixed} quantum system and the ensemble average that is supposed to capture its chaotic properties. In this work we pursue a set of parallel goals in order to investigate this question, with an eye on its holographic ramifications. Firstly we describe how the universal content of chaotic quantum systems is encapsulated in a simple symmetry-breaking principle, the {\it breaking of causal symmetry}. Secondly we show that this symmetry breaking principle allows one to select a universal set of light modes -- precisely the associated (pseudo--)Goldstone modes -- which quantitatively express the RMT physics, both for ensembles as well as individual chaotic quantum systems, in terms of a simple effective (field) theory, well-known in the chaos community as the Efetov-Wegner sigma-model~\cite{Wegner1979,efetov1983supersymmetry}. Thirdly, we argue that in the bulk causal symmetry is broken by a condensate of  strings, moving in the presence of D-branes representing the spectral determinants of the boundary theory. As an illustration we give an explicit bulk realization of this scenario in the context of minimal string theory, where the relevant effective theory is equivalent to Kontsevich's cubic matrix model.

The sigma-model approach is tailor-made to control the spectral statistics with a precision that is exponential in the size of the Hilbert space, i.e. sensitive to energy differences of the order
\begin{equation} \label{eq.EnergyResolution}
\Delta E \sim {\cal O}\left(e^{-S} \right)\,.   
\end{equation}
These contributions are
sometimes referred to as being doubly non-perturbative because they are controlled by factors $e^{-\mathrm{const.} \times e^S}$, exponential in the size of the Hilbert space,
which itself is exponential in the system size (e.g. $\sim e^{-e^N}$ for SYK on $N$
sites.) This is achieved by formulating the calculation of microscopically defined correlation functions
in terms of an effective theory of the  aforementioned pseudo-Goldstone modes. 
The large
parameter, dim${\cal H}$, stabilizes two saddle points on the Goldstone mode manifold, connected to
each other via a discrete Weyl group symmetry. Within this framework, structures such
as the `dip-ramp-plateau' region ubiquitous in spectral form factors are
 described as the contributions of non-ergodic fluctuations (dip), an
expansion in ergodic fluctuations around the first saddle (ramp), and one around the second  (plateau), respectively. The latter two structures are the universal content of the effective field theory of the Goldstone sector. The
Weyl group symmetry establishes a connection between the  expansions yielding ramp and plateau, respectively, and in this way gives access to deeply non-perturbative information (plateau) once the perturbative contents of a theory (dip/ramp) is under control. Finally, in the deep infrared, $\Delta E\lesssim {\cal O}( e^{-e^S})$,
the ergodic mode becomes massless, full integration over its nonlinear target manifold restores the causal symmetry, and in this way resolves the rigidity of level
repulsion in the chaotic many body spectrum in quantitative agreement with the results of random matrix theory (RMT).

In cases where a
theory is exactly given by a matrix model (say of matrix dimension $L\times L$), for
example in 2D gravity \cite{Ginsparg:1993is}, or the recent reformulation of JT gravity in terms
of matrix integrals \cite{Saad:2019lba},  the above approach improves on previous treatments in
that it gives analytical control over non-perturbative physics, i.e. contribution of
${\cal O}\left( e^{-L} \right)$. In situations where the theory is not exactly
equivalent to a matrix model, for example in the canonical holographic example of the
${\cal N}=4$ theory in $3+1$ dimensions, the sigma-model approach advocated in our
work gives a framework to understand in what sense random-matrix physics is a good
approximation, and to control spectral correlations non-perturbatively in terms of an
effective field theory.

The general idea of causal symmetry itself is quite simple: in order to extract
information about spectral correlations, it is necessary to consider correlation
functions between Green functions of opposite causality, that is to say between
advanced and retarded propagators. It is then convenient to develop a (path-)integral
representation of these correlation functions by introducing auxiliary degrees of
freedoms, at first separately for the advanced as well as the retarded sector.
However, the resulting system turns out to be invariant under continuous general
linear transformations which rotate advanced into retarded degrees of freedom (and
vice versa), the causal symmetry. Furthermore, in applications of interest to us,
this symmetry is broken both explicitly by a small amount and spontaneously by the
saddle-point solution(s).

 In such cases, one may describe the long time or low energy physics of the system in terms of an effective field theory assuming the symbolic form,
\begin{align}
\label{eq.SigmaModelEffectiveActionIntro}
    \mathcal{Z}=\int DQ \, e^{-S[Q]}.
\end{align}
Here, the integration is over fields, seen as maps from a base space $M$ to the
target space $T$, $Q:M\to T, x\mapsto Q(x)$, where $T$ is a low-dimensional nonlinear
target manifold universally determined by the symmetry breaking. For example, in many
cases of interest $Q(x)$ is simply a $4\times 4$ graded matrix, and $T$ the graded
coset space $\mathrm{U}(2|2)/\mathrm{U}(1|1)\times
\mathrm{U}(1|1)$ (hence the classification as  nonlinear sigma model.) By contrast,
the realization of the base space depends on the application. For example, in the
case of matrix models, the integral is over matrices $Q$, with no position dependence. In the classic applications of the theory to disordered electronic
systems, $Q(x)$ becomes a space-dependent matrix field, and $S[Q]$ an action
containing spatial gradient terms $ \int {\rm STr}(\nabla Q(x))^2$ (describing diffusive dynamics) in addition to
the explicit symmetry breaking mass term $\int  {\rm STr} (\omega Q)$. However, more relevant to holographic applications are many
body realizations of chaos, where $M$ reflects the
full structure of the underlying many-body Hilbert space. 
Given this potentially highly complicated structure it is fortunate that typically (that is in theories that do not exhibit many-body localization) there exists a threshold energy, the so-called Thouless energy, below which  fluctuations inhomogeneous on $M$ are energetically disfavoured, so that the field integral
collapses to one over a homogeneous `mean field' configuration $Q(x)=Q$. We thus conclude that, for these energies, the action reduces to that of the random  matrix models, thereby demonstrating RMT universality at low energies from the perspective of the EFT. The appearance of the scale $L$ in the exponent, which is a robust  construction principle of the present approach, means that the theory is `semiclassical' for energies $\Delta E\ll L^{-1}$, and capable of resolving `doubly non-perturbative' structures for $\Delta E \sim L^{-1}$. We remark that, as will become clear below,  the notion of `low-energy' in the EFT we develop here, refers to small {\rm differences} of energies in the spectrum of the original system.

The essence of the sigma model approach is a reduction of theories defined in
$e^S$-dimensional Hilbert spaces to effective theories on  much lower dimensional
`flavor' manifolds. Both the realization of the flavor manifold and its exact
dimensionality (which however remains always of $\mathcal{O}(1)$ in terms of $L$) depend on the symmetries of the
parent theory, and on the details of the correlation functions at hand, in a well-understood way. 

As we will reason later in the text, this reduction affords a beautifully simple bulk interpretation:
the flavor manifold  can be identified with the effective degrees of freedom of a
sector of open strings which stretch between a small number of `flavor branes', while
the pseudo-Goldstone modes are mesonic strings which begin and end on one of these
flavor branes. Again the nature and the number of `flavor branes' depends on the
observables one wants to compute, and is in exact correspondence with the
aforementioned flavor structure of the boundary theory. 

In the rest of the paper, we will discuss the contents prosaically summarized above
 in more mathematical language. Much of the paper addresses contents well familiar in
 either the string theory or the condensed matter/chaos community, but hardly any of
 it in both. We have tried to make the text accessible to readers from both camps,
 which explains a degree of redundancy and the presence of material which may be
 skipped by experts. In section~\ref{sec:SpectralProbes} we introduce 
 standard diagnostic observables probing the physics of chaotic systems, their
 representation in terms of graded field integrals, the principles of the above
 symmetry breaking mechanism, and that of the field theory defined by it. While
 the discussion of this section applies to a wide range of settings,
 section~\ref{sec:RMTSigmaModel} takes a complementary perspective and considers
 matrix theories --- with invariant yet general (not necessarily Gaussian)
 distribution --- as a concrete case study. This section intentionally goes into some
 technical detail. It can, but need not be read before the more exploratory
 section~\ref{sec.CausalSymmetryBreakingBulk}, where we address causal symmetry
 breaking from the perspective of the bulk. We present a discussion of some open issues Section \ref{sec.Discussion}, followed by Appendices going over salient details of impurity diagrams as well as the relation of the superymmetry technique used in the bulk of this paper and the replica approach.

\section{Spectral probes}
\label{sec:SpectralProbes}
\subsection{Time scales}
The prime focus of this work rests on late time chaotic properties, so it will
be useful to put the relevant scales in context. In order to do so, let us define a
few important quantities, starting with the density of energy levels
\begin{align}
\rho(E) = \frac{1}{L}\sum_{i=1}^L \delta \left( E - E_i \right)\,,
\end{align}
where $L = {\rm dim}{\cal H}$ is the dimension of Hilbert space. Since the quantum system of interest will have a very large number of such levels, it makes sense to speak of the typical separation at energy $E$, the mean level spacing
\begin{align}\label{eq.meanLevelSpacingands}
\Delta(E) := \langle \rho(E) \rangle^{-1}\,,
\end{align}
where the average denoted by angular brackets may be over energy windows of a given
Hamiltonian, a set of parameters, or as will often be the case, over an ensemble of
Hamiltonians. The mean level spacing defines the smallest energy scale in the problem, and its inverse, the \text{Heisenberg} time,  $t_H =2\pi\Delta^{-1}$ the
 largest time scale. (Referring to the emergence of a `plateau' in the form factor Eq.~\eqref{eq.DefSFF} of Gaussian unitary random matrices at this scale,  $t_H$ has been dubbed the `plateau time'.
 However,  given that in other symmetry classes there are no such signatures in the form factor, this may be a bit of a misnomer.) Following
 standard conventions in the field, we frequently measure energy differences
 $\omega=\mathcal{O}(\Delta)$ and their conjugate time scales $t$ in dimensionless
 variables
\begin{align}
     \label{eq.sTauDef}
     s\equiv \frac{\pi \omega}{\Delta},\qquad \tau \equiv \frac{t}{t_H}.
 \end{align} 
We  next outline some canonical time scales encountered in quantum chaotic systems,
in order to better define the regime of interest of this work, namely times of
parametric order $t_H$.

Firstly, let us generically assume that there is a parameter, such as 
$\hbar$ or $N$, so that the system is semi-classical for $\hbar, N^{-1} \rightarrow 0$.  A typical chaotic quantum system then has characteristic imprints resulting from
the following scales:
\subsubsection*{Early time chaos}
\begin{enumerate}
\item {\it Lyapunov time, $t_L = \lambda^{-1}$}: Many chaotic systems are defined with reference to a
semiclassical parameter $\hbar$ such that they become classical in the limit $\hbar
\to 0$. In these cases, the time scale $t_L$  characterizes the
exponential divergence $\delta x\sim e^{\lambda t}$ of initially close phase-space
trajectories.
In systems without straightforward classical limit, such as spin $1/2$ chains, time scales playing an analogous role may be defined in terms of observables showing early time exponential instabilities.
\item {\it Ehrenfest time} $t_E \sim \lambda^{-1} \log 1/\hbar$:  this quantum time
scale is attained when a minimum-uncertainty wavepacket localised to within a Planck
cell in phase space has spread to a width of order one. The prefactor of
$\lambda^{-1}$ is due to the semiclassical propagation of the wavepacket (i.e.
retains an imprint of classical Lyapunov chaos), while the factor $\hbar$ inside the
log comes from the quantum spread. This timescale is also known, especially in the
black-hole context, as the {\it scrambling time} $t_s$. In cases without
orthodox semiclassical parameter, $\hbar$, the scrambling time is reached when observables
probing early time instabilities have become `large', for instance at $t_E\sim \lambda^{-1}\ln(N)$. 
\item \textit{Ergodic time (aka Thouless time)} $t_E$: the time scale beyond which a
chaotic flow uniformly covers the classical energy shell in phase space. For example,
in a diffusive metal of linear extent $\lambda$ and diffusion constant $D$, this
would be the time $t_E\approx D/\lambda^2$ it takes to diffusively explore the
system. More generally, $t_E$ sets the scale  beyond which RMT behavior is attained. We caution that a Thouless time need not necessarily exist (such as in infinitely extended systems, or systems which do not quantum-thermalize in the long time limit), or can be be a tricky affair (such as in the SYK model, where  the fidelity to RMT depends on the observable under consideration).

\end{enumerate}
\subsubsection*{Late time chaos}
\begin{enumerate}
\setcounter{enumi}{3}
\item {\it Times exceeding the ergodic time $t> t_E$:} The physics described in this regime is what is often referred to as the `ramp' phase, associated with spectral rigidity of chaotic quantum systems. However, for $\tau\sim {\cal O}(1)$, necessarily non-perturbative physics sets in. This happens at the
\item {\it Heisenberg time} $t_{\rm H} = 2\pi\Delta^{-1}$: the behavior associated with this time scale is what is known as the `plateau' phase when speaking of certain observables, such as two-point functions or the spectral form factor. The physical behavior in this regime is non-perturbative in $\Delta$, but straightforward to  capture within the late-time effective description expanded upon in this work.
\end{enumerate}

To orient the reader, let us indicate these timescales on the example of the Majorana
SYK model \cite{Kitaev-talks:2015,Maldacena:2016hyu} on $N$ sites, which has Hilbert
space dimension $L=2^{N/2}$, dual to black holes in two dimensional anti-de Sitter
space \cite{Kitaev-talks:2015,Sachdev:2015efa}. A seminal computation by Kitaev
demonstrated that the scrambling time for this model takes the form $t_s =
\frac{\beta}{2\pi} \log(N)$, indicating that $N^{-1}$ assumes the role of the semiclassical parameter. The mean level spacing $\Delta \sim e^{-N}$ is exponentially small in $N$, the Heisenberg time scales like $t_H \sim
e^N$, and the `doubly non-perturbative' effects appearing when $\tau =1$ are of
order $e^{-e^N}$. These have been the object of much recent interest
\cite{Saad:2019lba}, and as part of this work we review how they are explicitly and analytically computable within the EFT approach\cite{Altland:2017eao}. This is in
contrast to the picture in \cite{Saad:2019lba}, where the existence of such effects
is inferred from the asymptotic behavior of a perturbative expansion in $\Delta$,
while the non-perturbative completion itself remains inaccessible.

\subsection{Resolvents and determinants}
Let us now introduce a set of observables well suited to characterise spectral properties of a quantum system of interest. We start by defining the trace of the resolvent
\begin{align}\label{eq.TraceResolvent}
W(z^\pm) = \left\langle{\rm Tr} \frac{1}{z^\pm-H} \right\rangle =  \left\langle{\rm Tr} G(z^\pm) \right\rangle\,,
\end{align}
where for the time being  the brackets indicate an average either over an energy
window or an ensemble of Hamiltonians and we have introduced the notation $z^\pm$
indicating the addition of a small positive or negative imaginary part the energy
argument. The utility of the spectral resolvent is in its simple relationship with
the spectral density, namely $\rho(z) = \mp\frac{1}{\pi} {\rm Im} W(z \pm i\delta)$.
Mostly we will be interested in higher spectral correlation functions, and
correspondingly in objects of the form
\begin{align}
W(z_1,z_2,\ldots z_k) =\left\langle{\rm Tr} G(z_1)  {\rm Tr} G(z_2)\cdots {\rm Tr} G(z_n)\right\rangle\,,
\end{align}
where each energy argument could have either a small positive or negative imaginary
part, and which shall often be indicated as $z^\pm_i$. A particularly interesting
such quantity is the spectral form factor, which is defined via the spectral
correlation function
\begin{align}\label{eq.SpectralCorrelationFunction}
R_2(\omega) = \Delta^2 \left\langle \rho(E + \frac{\omega}{2})\rho(E - \frac{\omega}{2}) \right\rangle_c\,,
\end{align}
(we denote the connected part $(\langle AB\rangle_c=\langle AB\rangle-\langle A\rangle\langle B\rangle)$ of any observable with subscript `$c$'), such that
\begin{align}\label{eq.DefSFF}
K(\tau) = \frac{1}{\Delta} \int d\omega R_2(\omega) e^{-i\frac{2\pi\omega}{\Delta} \tau}\,.
\end{align}
Note that the exponent is written in a natural way in terms of times normalized by the inverse level spacing.  Having discussed these characteristic probes, we now introduce a class of auxiliary quantities, which will turn out to be the most convenient objects on which we base our  framework. These take the form of ratios of determinants and we are principally interested in two cases, namely
\begin{align}\label{eq.SpectralDets}
{\cal Z}_{(2)}(\hat z) = \frac{{\rm det} (z_1 - H)}{{\rm det} (z_2 - H)}\qquad{\cal Z}_{(4)}(\hat z) = \frac{{\rm det} (z_1 - H){\rm det} (z_2 - H)}{{\rm det} (z_3 - H){\rm det} (z_4 - H)}\,.
\end{align}
By a slight abuse of notation, in each case the matrix $\hat z$ is taken to be the
diagonal matrix of ordered energy arguments, e.g. $\hat z = {\rm diag}(z_1,z_2)$ in
the first case. The dimension of this matrix will always be clear from the context,
and the utility of arranging energies in matrix form will emerge in our subsequent
developments. Let us briefly pause  and note that careful attention
needs to be placed on the infinitesimal imaginary parts given to the energy
arguments,  in the \emph{denominators}, where they determine the pole structure of the spectral determinants.  Specifically,  the spectral determinant ${\cal
Z}_{(2)}$ allows us to extract the spectral density via the resolvent, viz.
\begin{align}
\label{eq.Z2Intro}
W(z) = \partial_{z_2}{\cal Z}_{(2)}\left( \hat z \right) \Bigr|_{z_1 = z_2 = z} \quad \Leftrightarrow \quad \rho(z) = \mp \frac{1}{\pi L} {\rm Im} \,\partial_{z_2} \left\langle{\cal Z}_{(2)}\left( \hat z \right) \right\rangle \Bigr|_{z_1 = z_2 = E\pm i0}.
\end{align}
On the other hand, the spectral two point function requires the use of the ratio ${\cal Z}_{(4)}$: 
\begin{align}\label{eq.Z4Intro}
\left\langle \rho(E)\rho(E')\right\rangle = \frac{{\rm Re}}{\pi^2 L^2} \frac{\partial^2}{\partial z_3 \partial z_4} \Bigl\langle  {\cal Z}_{(4)}(z_1,z_2,z_3^+,z_4^-)   \Bigr\rangle \Biggr|_ {\substack{z_1 = z_3^+=E\,\\ z_2 = z_4^-=E'  }}.
\end{align}
The spectral determinant $\mathcal{Z}_{(4)}$ possesses an interesting {\it Weyl symmetry}  under the exchange of $z_1 \leftrightarrow z_2$, which leaves the spectral correlation function unchanged. As we will discuss later, this discrete symmetry is key to the understanding of the non-perturbative structure of the spectrum. It is called  'Weyl symmetry' because in the later representation
of the problem as a coset matrix integral the reflection $z_1 \leftrightarrow z_2$
translates to a Weyl group symmetry in the mathematical sense on the symmetry group
of the integral. 

\subsection{Causal symmetry}
A very useful way to rewrite the ratios \eqref{eq.SpectralDets} is by means of a graded Gaussian integral
\begin{align}\label{eq.GaussianGradedGenFun}
{\cal Z}_{(4)}\left( \hat z\right) = \int e^{-i \bar\psi (\hat z - H)\psi}d(\psi ,\bar\psi)\,,
\end{align}
where $\psi \in \cV$ is a $4L$ dimensional graded vector with $2L$ Grassmannian
components and $2L$ ordinary c-number components. Here, the Grassman integrals
produce the determinant factors in the numerator, while the c-number integrals
produce those in the denominator. It should also be clear how to write the
corresponding expression of a ratio ${\cal Z}_{(2n)}$ in terms of a $2nL$-dimensional
graded integral, although we shall not be needing the higher $n>2$ cases in the
present work. The detailed structure of the graded vector space is associated to
different physical aspects of our problem and takes the form $\cV = \cH \otimes
\cV_F$, where $\cH$ is the Hilbert space our Hamiltonian $H$ acts on, and we refer to
$\cV_F$ as ``flavor space". We note that in this context the Hilbert space
itself is often referred to as ``color space''. We will often use this terminology in this work, but caution the reader not to confuse this usage of ``color" with the common usage of color symmetry in Yang-Mills theory.  On the other hand, the flavor-structure is dictated
by the requirement that each determinant must come with an inverse determinant so as to
ensure normalization of the final physical quantities. This introduces a
$\mathbb{Z}_2$ grading to flavor space\footnote{Let us note that the normalization of
final results can also be achieved by means of the replica trick, which would lead to
a much bigger, but purely bosonic flavor space. We shall comment on this alternative
point of view from time to time.}. Finally, in the cases of interest we have advanced
and retarded sectors which each add one more factor to the tensor product making up
flavor space.

 Let us now describe more closely the structure of the integral \eqref{eq.GaussianGradedGenFun} above: as mentioned, each of the fields $\psi=\{\psi^a_\mu\}$ carries a Hilbert-space index $\mu$ as well as a flavor index $a$, suppressed above for notational transparency. The action of the Gaussian integral is subject to a  $\mathrm{GL}(2L|2L)$ symmetry in the fundamental representation, 
\begin{align}
\label{eq.PsiTransformationFullGroup}
 \mathrm{GL}(2L|2L) \qquad \Rightarrow \qquad \psi \to g \psi\,,\quad  \bar\psi \to \bar\psi g^{-1}\,,
\end{align}
 weakly broken by the energy matrix $\hat z$, and strongly broken by the Hamiltonian $H$. (We here use standard $(\cdot|\cdot)$ notation in referring to group representations in graded spaces.)  The origins of this symmetry and its (spontaneous) breaking in the causal sector will be the main guiding principle in the construction of this paper. Secondly, the index $a$ plays a double role. It labels both fermionic and bosonic components, as well as advanced and retarded components, distinguished by their respective imaginary offsets. One may thus also write $a= (\sigma,s)$, where $s=\pm$ labels the component in the advanced-retarded basis and $\sigma$ denotes the fermion-boson grading. Thirdly, the action of conjugation (see section \ref{sec:RMTSigmaModel}, after Eq. \eqref{eq:SpectralDeterminantGaussian} for the detailed definition) $\bar\psi$ together with the infinitesimal imaginary offsets $\delta_a$ ensure convergence in the bosonic sector.

The  field theory approach to quantum chaos takes the exact
representation~\eqref{eq.GaussianGradedGenFun} as the starting point towards the
construction of an effective low energy theory describing the system at large time
scales. Here `low energy' refers energy differences of order $|z_i-z_j|\sim  \Delta$,
assumed to be of the order of the quantum energy level spacing
\eqref{eq.EnergyResolution}. In certain examples, these theories can be derived from
first principles where both the final form of the effective action, and the details
of the construction are specific to the physical system at hand. However, we
here reason from the vantage point of symmetries,  which lets the underlying
structures stand out, and defines generally applicable construction principles. Specifically,
symmetries  determine the target spaces of the effective field theories, the
essential strategy of their derivation, their operator contents, and the universal
physics of the ergodic phase (where one exists.) In the following, we introduce
these symmetries and their manifestations in the effective low energy theory from a
birds eye perspective. This discussion is complemented in
section~\ref{sec:RMTSigmaModel} by an exemplary derivation of the
effective field theory for the simple case where $H$ is drawn from a 
 matrix ensemble. In section~\ref{sec.CausalSymmetryBreakingBulk} we give an explicit bulk
realization, again for what is arguably one of the simplest examples, namely $(2,q)$
minimal string theory, and draw broader conclusions for
holographic duality.

\subsection{Ensemble vs. individual system}
{\it Unus pro omnibus, omnes pro uno}

{\hspace{2cm}\small (The unofficial motto of Switzerland)}
\vskip1em
\noindent We have at various points stated that causal symmetry breaking is the
universal description of quantum chaotic systems. However, this begs the question to
what extent individual chaotic systems display this `universal' behavior. Our
approach to quantum chaos is inherently statistical, describing systems in terms of
correlation functions which either directly or effectively make reference to
(parametric) ensembles. For an individual system the notion of statistics does not exist in a strict sense\footnote{unless one admits the  distribution of its energy levels
along the real axis as a discrete measure --- an approach that worked miraculously
well since the early days of the field when the resonance spectra of individual heavy
nuclei were put in relation to RMT spectral statistics.}. On the other hand we know that if we average over an ensemble of microscopically different but macroscopically identical systems, universal statistics emerges. On this basis, the expectation  is
that in spectral correlation functions computed for individual chaotic systems,
smooth backgrounds exhibiting RMT behavior appear superimposed with high frequency
noise --- see Ref.~\cite{Braun_2015} for a case study demonstrating this phenomenon. For a system of Hilbert space
dimension $L$, the noise amplitude scales with $\sim L^{-1/2}$, masking the rapid
universal decay of spectral correlations already for small energy differences.
However, these fluctuations are extremely rapid, making them susceptible to dephasing
under any mild averaging. This has the effect that signal smoothing by \textit{any}
continuous averaging protocol, over external system parameters, 'disorder', or even
the value of $\hbar$, efficiently eliminates it and allows the universal
content to emerge.  These observations comport well with recent work that proposes to
define statistical ensembles for AdS$_3$ gravity by averaging over a set of moduli of
the compactification \cite{Afkhami-Jeddi:2020ezh,Perez:2020klz,Maloney:2020nni,Cotler:2020ugk}, which serve as the
small set of smoothing parameters necessary to bring out the EFT behavior.

If one wants to strictly confront individual chaotic systems, without any mild
averaging over a small set of parameters whatsoever, the EFT can be stabilized by an
average over
energy~\cite{AltshulerAndreevSimons,AltlandMicklitz,AltlandGnutzmannHaakeMicklitz}. A
subsequent expansion in smooth field fluctuations then yields an action which, by
design, disposes with the information on high frequency noise. Its capacity to yield
universal information beyond the RMT limit has been demonstrated on a number of case
studies, including the field theory approach to quantum
graphs~\cite{AltlandGnutzmann}, or to nonperturbative localization phenomena in the
quantum standard map~\cite{Tian_2010}. While slightly going against our EFT logic, it
may be instructive to apply such an approach to one of our canon of field theories
with holographic duals in higher dimensions, e.g. ABJM theory or the ${\cal N}=4$ SYM
theory.

\subsection{The effective field theory of quantum chaos}\label{sec.EFTofChaos}
Having qualitatively outlined the main ideas going into developing an effective theory of universal spectral correlations in quantum chaotic systems, we now delve into the conceptual steps involved in its construction in some more detail.
\begin{enumerate}
    \item {\it Broken symmetry and the analytic structure of the resolvent:} consider the resolvent $G(z)\equiv (z-H)^{-1}$ of a chaotic Hamiltonian, where we will assume that $L = {\rm dim}{\cal H} \gg 1$. For example, considering the case where $H=H(v)$  depends on randomness (symbolically represented by the variable $v$) and  the averaging in Equation~\eqref{eq.TraceResolvent} is over a distribution $P(v)$ of the latter. Prior to  averaging, the resolvent has poles at the discrete eigenvalues of $H(v)$, implying that  $\mathrm{Im}\,\mathrm{tr}(G(z))=0$ almost everywhere for $z=\epsilon\pm i0$. By contrast, the average resolvent $\langle G(z)\rangle$ has a branch cut inside the spectral support of $H$ along the real axis, where $\mathrm{Im}\, \mathrm{tr}\langle G(z)\rangle=\pm \mathcal{O}(\rho(z))$, the sign being uniquely determined by that of the infinitesimal imaginary part of the energy argument, $\mathrm{Im}(z)$. 
    This illustrates the simplest instance of a symmetry breaking scenario characterized by an amplification $\pm i0\to \pm i\rho$ of the infinitesimal imaginary energy increments to a large and finite value, proportional to the averaged spectral density, $\rho(z)$.
    \item  {\it Pattern of symmetry breaking:} the consequences of the symmetry
    breaking in the effective theory become apparent when we discuss it in the
    context of the transformation group Eq.~\eqref{eq.PsiTransformationFullGroup}.
    Since $H$ represents an explicit and strong symmetry breaking in the `color'
    sector, only transformations in $\mathrm{GL}(2L|2L)\to \mathrm{GL}(2|2)\times
    \mathds{1}_L\equiv \mathrm{GL}(2|2)$ can be symmetries
    of the late time physics.. The above breaking mechanism collapses this
    flavor symmetry group to $\mathrm{GL}(1|1)\times
    \mathrm{GL}(1|1)$ --- two dimensional transformations between bosonic and
    fermionic degrees of freedom acting separately in the sector of retarded and
    advanced indices, $s=\pm$. The transformation between the two causal sectors are
    thus spontaneously broken, whence the term `causal symmetry breaking'.
    Furthermore the symmetry is explicitly, but weakly broken by the differences in
    the energy arguments entering the diagonal matrix $\hat z$. We thus  conclude
    that the degrees of freedom essential to the low energy physics of the system are
    flavor Goldstone modes drawn from the  manifold\footnote{The presence of
    anti-unitary symmetries in the microscopic theory, such as time reversal, charge
    conjugation, or chiral symmetries, restricts the set of symmetry-compatible
    continuous transformations and changes the Goldstone mode manifold.
    However, for simplicity, we focus on the simplest setting, where $H$ is just
    hermitian.}
\begin{align} \label{eq.GoldstoneCoset}
 {\cal M} =  \mathrm{GL}(2|2)/ \left(\mathrm{GL}(1|1)\times \mathrm{GL}(1|1) \right)\,.
\end{align}
    \item {\it Goldstone modes:} since the physics is effectively dominated by the light modes, the  effective theory will be given by an integral over the soft manifold. The convergence of this integral requires a reduction to the coset space $\mathrm{U}(2|2)/\mathrm{U}(1|1)\times \mathrm{U}(1|1)$, where the bosonic (fermionic) sector of $U(2|2)$ is the (pseudo)unitary group in two dimensions. Geometrically, the bosonic sector of the coset is a two-dimensional hyperboloid, and the fermionic one a two-dimensional sphere.  A convenient way to represent this manifold is in terms of the matrix field
 \begin{align}\label{eq.ExponentiatedPions}
 Q=T\tau_3 T^{-1}\,, \qquad T\in \mathrm{U}(2|2)
 \end{align}
  where $\tau_3$ is the Pauli matrix in causal space $(\tau_3)^{ss'}=(-)^s
  \delta^{ss'}$, and the action of $T$ by conjugation respects the residual
  $\mathrm{U}(1|1)\times \mathrm{U}(1|1)$ symmetry. More generally, the Goldstone
  modes $Q(x)$ are fluctuating degrees of freedom, where $x$ parameterizes the
  base space of the effective theory. In this case, the Goldstone modes are
  constructed by writing $T$ as a product of an element in the full group $G$ times
  an element of the preserved subgroup $H$. This is achieved by letting $T(x)\to
  T(x)H(x)$, $H(x)\in \mathrm{U}(1|1)\times \mathrm{U}(1|1)$ so that $H$ becomes an
  exact local symmetry, and $T(x)\to T_0 T(x)$, $T_0\in \mathrm{U}(2|2)$ a global
  one, weakly broken by the differences of the energy arguments entering the argument
  $\hat z$. 
 
    \item {\it Effective action:} Historically, the present form of nonlinear sigma models  systems was pioneered in their  application  to the physics of disordered metals. There, $x$ is a real space coordinate, and the action assumes the form     
   \begin{align}\label{eq.EFTactionInTermsofvF}
    S[Q] = -iv \int {\rm STr} \left( \hat z Q \right) dV+ F^2 \int {\rm STr} \left( \nabla_i Q \nabla^i Q \right) dV + \cdots ,
    \end{align}
where   `$\mathrm{STr}$' is the matrix trace generalized  to `supermatrices' containing commuting and anti-commuting elements.\footnote{For a supermatrix $F =\left( \begin{smallmatrix}
a & \sigma \\ 
\tau & b
\end{smallmatrix}\right) $ with bosonic blocks $a,b$ and fermionc blocks
 $\sigma,\tau$ is defined as $\,\mathrm{STr}\,F = {\rm Tr}\,a - {\rm Tr}\,b$.}
The dimensionful quantities $v$ and $F^2$ can be identified (see section \ref{sec:RMTSigmaModel}) with the density of states per volume, $\rho(z)$, and the diffusion constant $D$ respectively.   Note
 the structural similarity to the chiral Lagrangian of QCD \cite{Weinberg:1966kf}
 (albeit with a different symmetry-breaking pattern) which shows that the diffusion
 term enters in the same way into the effective field theory of chaos as the Pion
 decay constant enters into the chiral effective Lagrangian\footnote{One can write
 down higher order terms in the symmetry breaking parameter such as $\str(Q\hat z
 Q\hat z)$ by promoting $\hat z$ to a spurion field. Higher powers of $\hat z$ are
 then suppressed by  powers of $v/F^4$ leaving us with having to deal only with the
 most relevant one we wrote.}. However, unlike the QCD Lagrangian, $S[Q]$ does not describe a dynamical theory, the formal reason being that we consider correlations at fixed energy, so that time is effectively frozen out.

 However, most relevant to the present context are recent extensions of the formalism to interacting systems such as the SYK model\cite{PhysRevLett.118.127202,Altland:2017eao,monteiro2020fock}. In these cases, the  coordinates $x\to n$ label the $L \sim e^{S}$ discrete basis states of the many-body Hilbert space, and the action assumes the symbolic form 
 \begin{align}\label{eq.EFTactionFockSpace}
    S[Q] = -i v \sum_n {\rm STr} \left( \hat z Q \right)+ \sum_{n,m} W_{nm}  {\rm STr} \left(Q_n  Q_m \right) + \cdots ,
    \end{align}
    where $v$ is now the density of states per lattice site, and $W_{nm}$ are the
    lattice hopping matrix elements defined by the underlying many body Hamiltonian.
    (For example, the four Majorana interaction defining the SYK
    Hamiltonian~\cite{Kitaev-talks:2015} can change up to four fermion occupation
    numbers, making for a range four hopping operator $W_{nm}$.) Depending on the strength of the hopping elements, the above model describes a thermalizing phase where below a finite Thouless energy the uniform mode $Q_n\to Q$ dominates (this happens, e.g., in the application to the SYK model), or a `many body localized' phase with strong independent fluctuations of $Q_n$ (see Ref.~\cite{Evers2008} for review.)

    In ergodic regimes, both the action describing single particle systems Eq.~\eqref{eq.EFTactionInTermsofvF}, and many body systems, Eq.~\eqref{eq.EFTactionFockSpace} collapse to the zero mode action 
\begin{align}
    \label{eq.SErgodic}
    S(Q)=-i \frac{\pi}{2\Delta}\mathrm{STr}(Q\hat z),
\end{align}
where $Q(x)=Q_n=Q$ is the homogeneous zero mode and the level spacing defined as $\Delta^{-1} =\frac{2}{\pi}v V = \frac{2}{\pi}v \sum_n 1$ defined through the effective dimension of Hilbert space. In this regime, the systems are physically equivalent to random matrix models (which are likewise described by the action~\eqref{eq.SErgodic} as we will demonstrate in section~\ref{sec:RMTSigmaModel}):

\item {\it Extracting the random-matrix physics:} The partition sum describing a chaotic quantum system in the ergodic regime assumes the form
\begin{align}
    \label{eq:QMatrixIntegral}
    \langle\mathcal{Z}(\hat z)\rangle=\int dQ\,e^{-S(Q)}\,,
\end{align}
where the action is given by Eq.~\eqref{eq.SErgodic} and the integral is over a
single instance of the matrices Eq.~\eqref{eq.ExponentiatedPions}. Thinking of $Q$ as
elements of a generalized sphere, and $dQ$ the corresponding  invariant measure
(section~\ref{sec:RMTSigmaModel} will provide the details), the action is that of a
`magnetic field' of strength $\sim s=\pi \omega/\Delta$. That action comes with two
saddle points, a stable one on the north pole with action $0$, and an unstable one at
the south pole with action $2is$. The fluctuations around both poles are suppressed in
the same parameter $s$. In this way, we can understand how the present formalism
produces Eq.~\eqref{eq.SpectralTwoPointUnitary} for the spectral function. In the
next section, we add some physical contents to the mathematical structure of
Eq.~\eqref{eq:QMatrixIntegral}. We will discuss the semiclassical interpretation
around the two saddles, its connection to the ramp-plateau profile of spectral
correlations, and the  idea of a holographic bulk interpretation of these
structures.

\end{enumerate}

 \subsection{The ergodic sector of the EFT and its topological expansion\label{sec.TopologicaExpansionEFT}}
 In most holographic applications to date, one is interested in systems which have an ergodic limit (i.e. not many-body localized) and in this work in particular, we are interested in the universal behavior of the ergodic limit. 
\subsubsection*{Saddle points and Weyl symmetry} 
 Let us thus take a closer look at the EFT in its ergodic phase, i.e. the integral Eq.~\eqref{eq:QMatrixIntegral}, starting at first in
 the case $s\gtrsim 1$ of energy splittings exceeding the level spacing. In this
 case, the integral will be dominated by small fluctuations around its stationary points, $\bar
 Q$ where the latter are identified by  stationarity under variations,
 $\delta S[\bar Q]=0$. Using the representation Eq.~\eqref{eq.ExponentiatedPions}, it
 is straightforward to verify that the stationarity condition is equivalent to $[\bar
 Q,\hat z]=0$, or matrix-diagonality of $\bar Q$. A somewhat closer analysis shows that of the four saddle points
 compatible with the unit-modularity of the eigenvalues $\text{spec}(Q)=\{1,1,-1,-1\}$ only
 two are compatible with the manifold structure.

 To understand this in an intuitive way, we recall that the commuting variables
contained in $Q$ parameterize $H^2\times S^2$, the product of a (non-compact) hyperboloid and a
(compact) two sphere. We can conveniently parametrize the compact sector $Q^{\mathrm {ff}}=T^{\mathrm{ff}}\tau_3
T^{\mathrm{ff}-1}=n_i \tau_i$ in terms of the unit vector, $\mathbf{n}$. Then the
two saddle point are the \emph{standard saddle}, $ Q_0=\text{bdiag}(\tau_3, \tau_3)$, and the
\emph{Altshuler-Andreev saddle}~\cite{AltshulerAndreev} $Q_\mathrm{AA}\equiv
\mathrm{diag}(\tau^{\mathrm{bb}}_3,-\tau^{\mathrm{ff}}_3)$.  In terms of the previously defined unit vector, we see that the standard saddle is located at the north pole of the sphere $\mathbf{n} = (0,0,1)$ and the Altshuler-Andreev saddle is at the south pole, $\mathbf{n} = (0,0,-1)$. Noting that the spectral probes of interest, \eqref{eq.SpectralDets}, are computed in the particular configuration $\mathrm{Re}\,z_1=-\mathrm{Re}\,z_2=\mathrm{Re}\,z_3=-\mathrm{Re}\,z_4=-\omega/2$, we
obtain the corresponding actions as $S[Q_0]=0$ and  $S[Q_\mathrm{AA}]=-2is$,
respectively.

Finally, notice that the map $Q_0 \to T_W Q_0 T_W^{-1}\equiv Q_\mathrm{AA}$ permuting
diagonal matrix elements is an element of the Weyl group of the underlying supergroup
structure. The above operation is equivalently described by a transformation of
energy arguments, $T_W^{-1}\hat z T_W\equiv \hat z'$, which  permutes
$z_1\leftrightarrow z_2$, an exchange introduced in connection
with~\eqref{eq.Z4Intro} as a Weyl symmetry of the spectral determinant. Writing the
Weyl symmetry transformation as
 \be
 \str(Q_\mathrm{AA}\hat z)=\str(Q_0 \hat z')\,,
 \ee
  we can see that the Weyl exchange of the energy arguments transforms the stationary configuration from the standard to the AA saddle. This raises the possibility that one can `bootstrap' the non-perturbative content of the Altshuler-Andreev saddle, by using the Weyl group transformation on the theory written around the standard saddle. Indeed,  in connection with periodic orbit theory~\cite{Muller2005} permutations of energies in the spectral determinant have been applied to access non-perturbative information from perturbative orbit expansions. (See remarks below Eq.~\eqref{eq.SpectralTwoPointUnitary} for the connection of this operation to the Riemann-Siegel lookalike hypothesis~\cite{Keating92} for the extension of semiclassical analysis.)

In the following we take a closer look at the contribution of these two stationary
points to spectral correlation function sand their interpretation in a holographic bulk
language.

\subsubsection*{Perturbative contributions: wormholes and baby universes}
Before delving into the bulk story in section~\ref{sec.CausalSymmetryBreakingBulk}, let us explain qualitatively how the
formalism naturally produces correlations of the type associated with Euclidean
wormholes. As is well known from the study of low-energy QCD --- or indeed any other
context where pseudo-Goldstone bosons dominate the physics --- the physical
manifestation of the coset are the Goldstone modes (aka pions) associated to the
generators of the broken symmetry. Let us schematically write
\begin{align}
    \label{eq.BandWmatrices}
    T=\exp(W),\qquad W=\left(\begin{matrix}
        &B\cr \tilde B &
    \end{matrix}  \right),
\end{align}
where $W$ is
the matrix of pion fields, expanded in the broken generators $B$ and $\tilde B$, which individually are $(1|1)$ supermatrices. To leading order in these generators,  the action takes the
form $S[B, \tilde B] = -2i s \,{\rm STr} \left(B\tilde B \right)$,  where for the time being we expand around the `standard saddle'. We have written the
exponent to leading (quadratic) order only, which is justified by the EFT logic and
which we will correct systematically. Expanding around this quadratic limit, we are led to consider matrix integral averages of the type
\begin{align}\label{eq.BAverage}
    \langle \dots \rangle \equiv \int d(B,\tilde B) \, e^{-2is\,{\rm STr} \left(  B \tilde B  \right)}(\dots)\,,
\end{align}
where the operator insertions are similarly built up from the $B,\tilde B$ matrices.
Specifically, in the computation of the two-point function at leading order we are
instructed to compute the expectation value $\langle  {\rm STr} ( B\tilde B
{\color{red}P^\mathrm{f}})  {\rm STr} ( \tilde B B {\color{blue}P^\mathrm{f}} )
\rangle $, where the matrices
$P^\mathrm{f}=\{\delta_{\sigma,\mathrm{f}}\delta_{\sigma',\mathrm{f}}\}$ project onto
the Grassmann sector.  The matrices $B$ and $\tilde B$ each have one index acting in
the retarded sector and one index acting in the advanced sector, and the order of
multiplication in the preceding expectation value effectively projects once on the
advanced (red) and once on the retarded sector (blue), as required by the
prescription in \eqref{eq.SpectralDets}. We may then represent the matrices $B,
\tilde B$ using double line notation, the Wick contraction of these matrices with the
Gaussian weight Eq.~\eqref{eq.BAverage} can be represented as
 \begin{align}\label{eq.GoldstoneCylinder}
\includegraphics[scale=0.35]{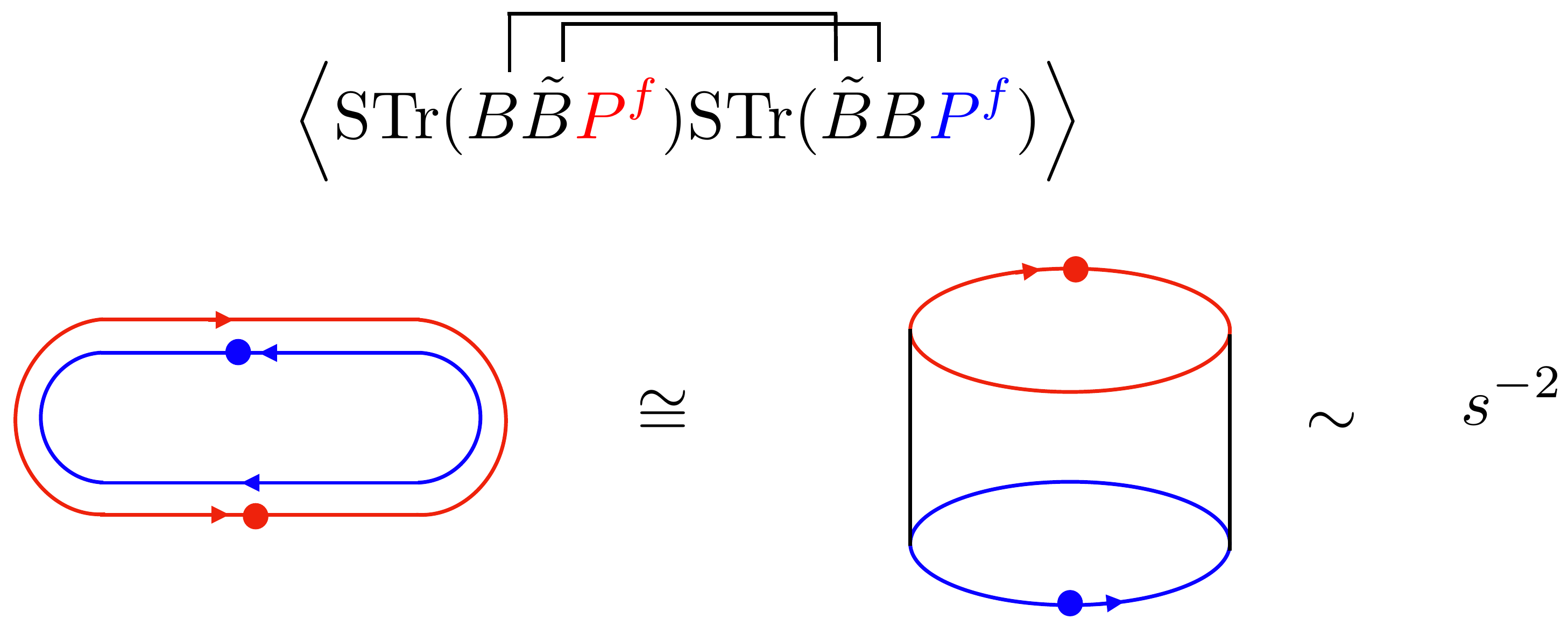}
 \end{align}
 where the marked (red) points are the insertions of the projectors, which appear on
 the boundaries of the cylindrical surface, as indicated. (Readers
 interested in a more microscopically resolved interpretation of the above
 correspondence are invited to read
 Appendix~\ref{ssub:expansion_in_goldstone_modes_vs_topological_recursion}.) By
 viewing the matrix contractions as defining a Riemann surface via their ribbon
 graph, one quickly convinces oneself that the Wick contraction as shown gives rise
 to an annulus (or cylinder) type contribution. These are associated with Euclidean
 wormhole type geometries in the bulk
 \cite{Saad:2019lba,Stanford:2019vob,Marolf:2020xie,Maldacena:2004rf}, as we will demonstrate directly in
 Section \ref{sec.CausalSymmetryBreakingBulk} below. In this way our formalism
 automatically produces connected correlations between different determinants,
 preventing their expectation values from factorizing.  The two Wick contractions
 give one factor of $s^{-1}$ each, resulting in a total contribution $\sim s^{-2}$.
 Fourier transformed into the time domain, this gives precisely the linear in time
 behavior characteristic of the ramp. It has often been pointed out (see e.g.\cite{Saad:2018bqo}) that the ramp physics is intrinsically non-perturbative, which is seen from our perspective from the fact that $s^{-2} \sim e^{-2S}$. The sigma-model approach transforms such -- in principle --non-perturbative contributions into a simpler perturbative expansion (around the standard saddle).
 
  However, as should be clear from the relation $Q = e^W\tau_3 e^{-W}$ this is but the leading order contribution in the expansion, and we now move on to higher-order examples. The next simplest diagram comes from including in the action the term proportional to ${\rm STr}(B\tilde B)^2$ in the expansion of $Q$, which gives us a contribution
  \begin{align}\label{eq.GOEcrosscap}
\includegraphics[scale=0.35]{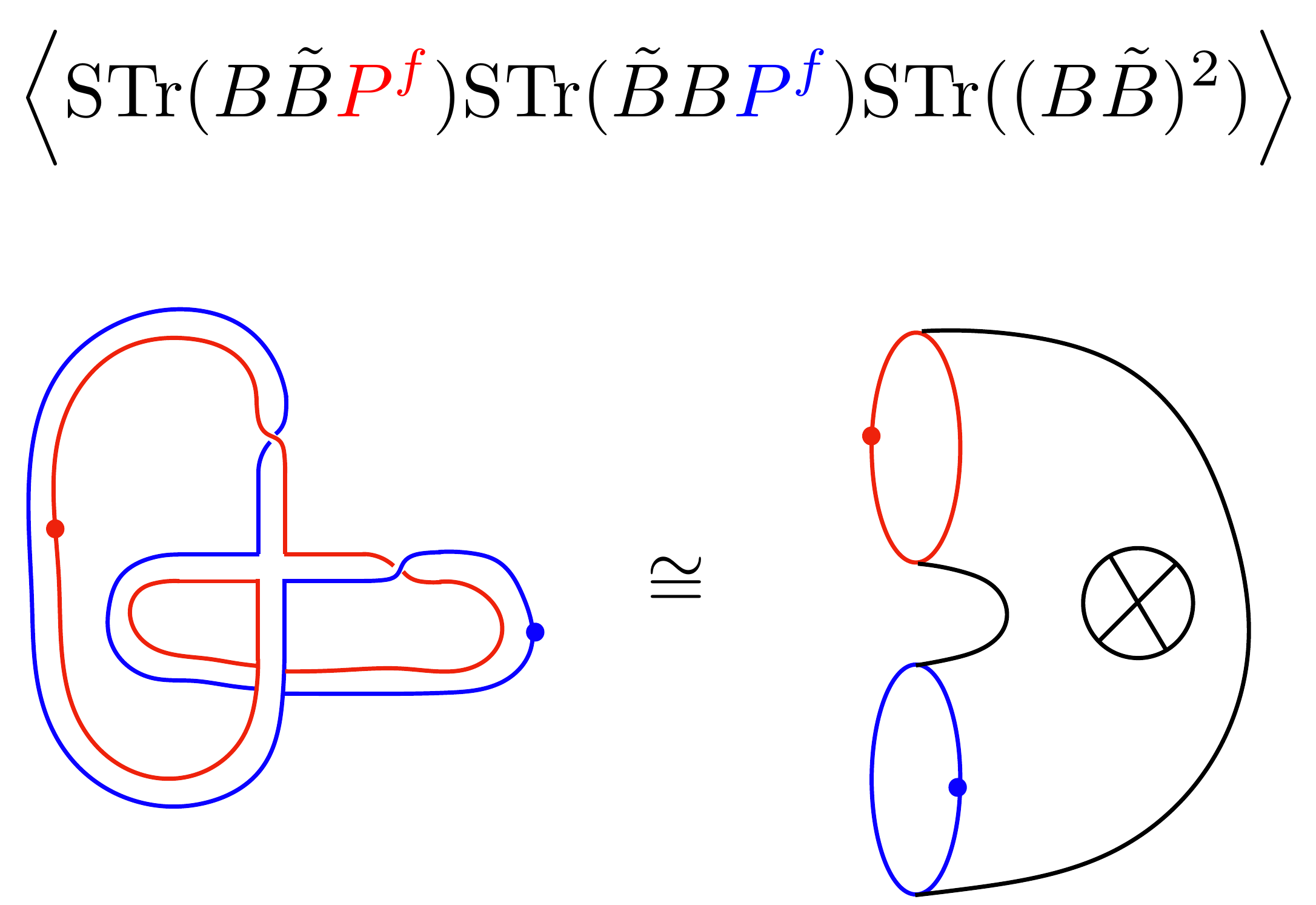}
 \end{align}
 To avoid clutter, we have not shown the Wick contraction which, however, is reflected in the structure of the ribbon graph itself. We note that this contribution can only appear in a theory that has time reversal invariance, for example in the GOE symmetry class, which allows un-oriented ribbon graphs. The reader may convince herself that there is in fact no consistent set of arrows that can be drawn on the two closed loops of the ribbon graph above, so that each each is always anti-parallel to its neighboring edge (again, we refer to Appendix~\ref{ssub:expansion_in_goldstone_modes_vs_topological_recursion} for a more microscopically resolved discussion of this point.) Correspondingly, the bulk surface is non-orientable which is achieved  by the crosscap insertion, as realized in bulk language in \cite{Stanford:2019vob}. Finally, by expanding to yet higher order in the $B,\tilde B$ matrices, we can find higher-genus contributions, the first non-trivial one coming from expanding to order str$((B\tilde B)^4)$. This adds one handle as in
   \begin{align}
   \label{eq.TwoCrossesOrientable}
\includegraphics[scale=0.35]{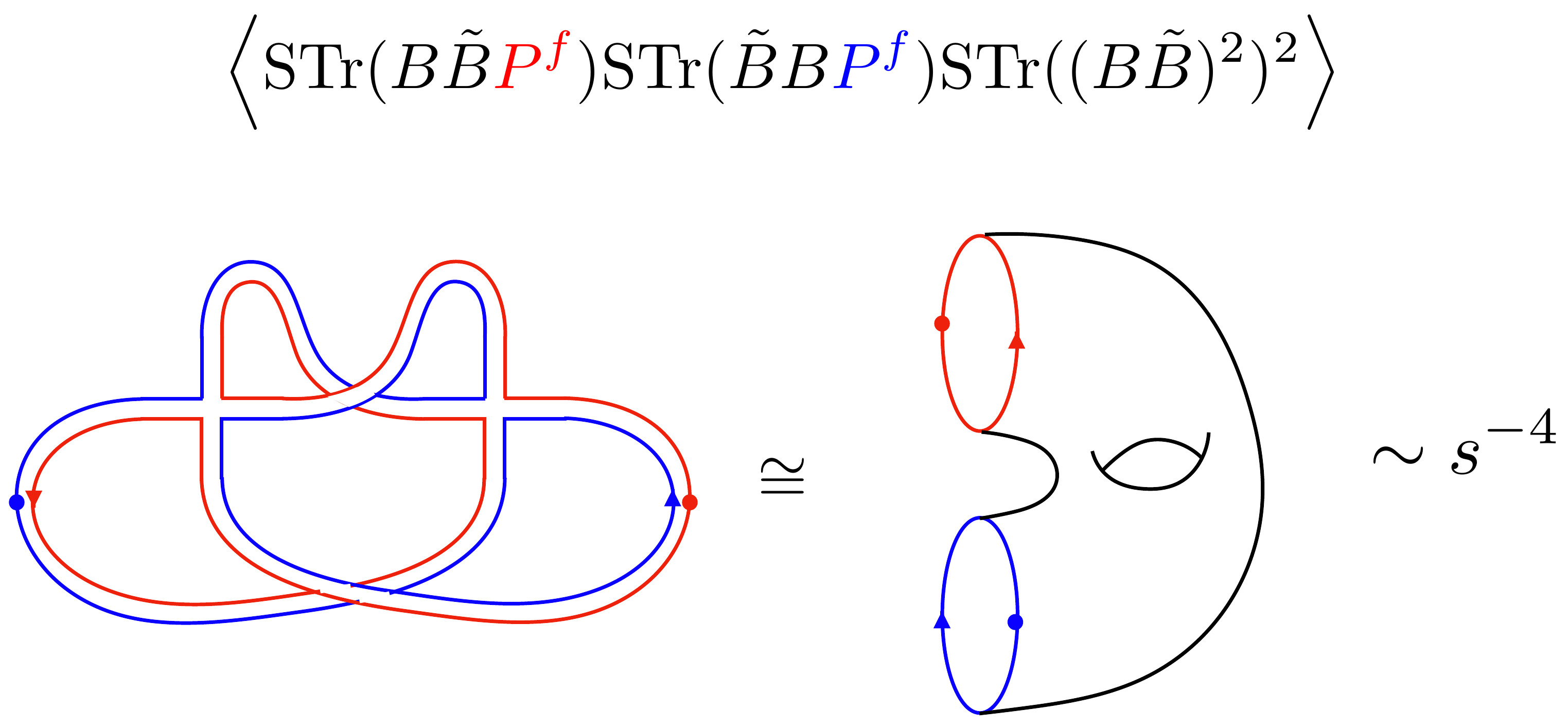}
 \end{align}
In the unitary class, it actually turns out that a further genus-one contraction exists, which cancels precisely against the one shown here, but in other symmetry classes diagrams like the one above give non-vanishing contributions. Since the translation between ribbon graph and surface by eye can become a little complicated, one can proceed more abstractly. By tracing the loops of this diagram one sees that it is indeed orientable and that it has $F= 2$ faces, $V=2$ vertices and $P=4$ propagators, meaning that it has a single handle $H=1$. This follows directly from the classic formula in toplogy  $F-P+V = 2 - 2H$. The marked points corresponding to the $P^\pm$ projectors are again interpreted as brane boundaries. The bulk interpretation is a spacetime with two boundaries and non-trivial topology, i.e. a case where a baby universe has split off and re-fused in an intermediate channel.

The full perturbative expansion of the EFT  proceeds by including higher and higher terms in the expansion of $Q$ in terms of the $B$ matrices. This results in successively higher-genus surfaces, each with two holes. The number of such holes is fixed to two, since we are computing a spectral two-point function. It should be clear how this generalizes to higher-point functions. 
  Note that the mathematical structure necessary for a connected correlator
  between different spectral determinants is a robust consequence of the
  symmetry-breaking scenario and its associated Goldstone physics.
  Notably this also gives a well-defined meaning to such correlations in a theory
  with fixed chaotic Hamiltonian and gives  legitimacy to the appearance of Euclidean bulk
  wormholes in such cases. 

  Finally we should note that the expansion here has some
  similarity with the topological expansions of the JT matrix model
  \cite{Saad:2019lba} or indeed that of the older matrix models \cite{Ginsparg:1993is}. We will have more to say about this in Section \ref{sec.CausalSymmetryBreakingBulk}. 

\subsubsection{Non-perturbative structure: second saddle and symmetry restoration}
\label{sec.SecondSaddle}

\begin{figure}[t]
\begin{center}
\includegraphics[width=.6\columnwidth]{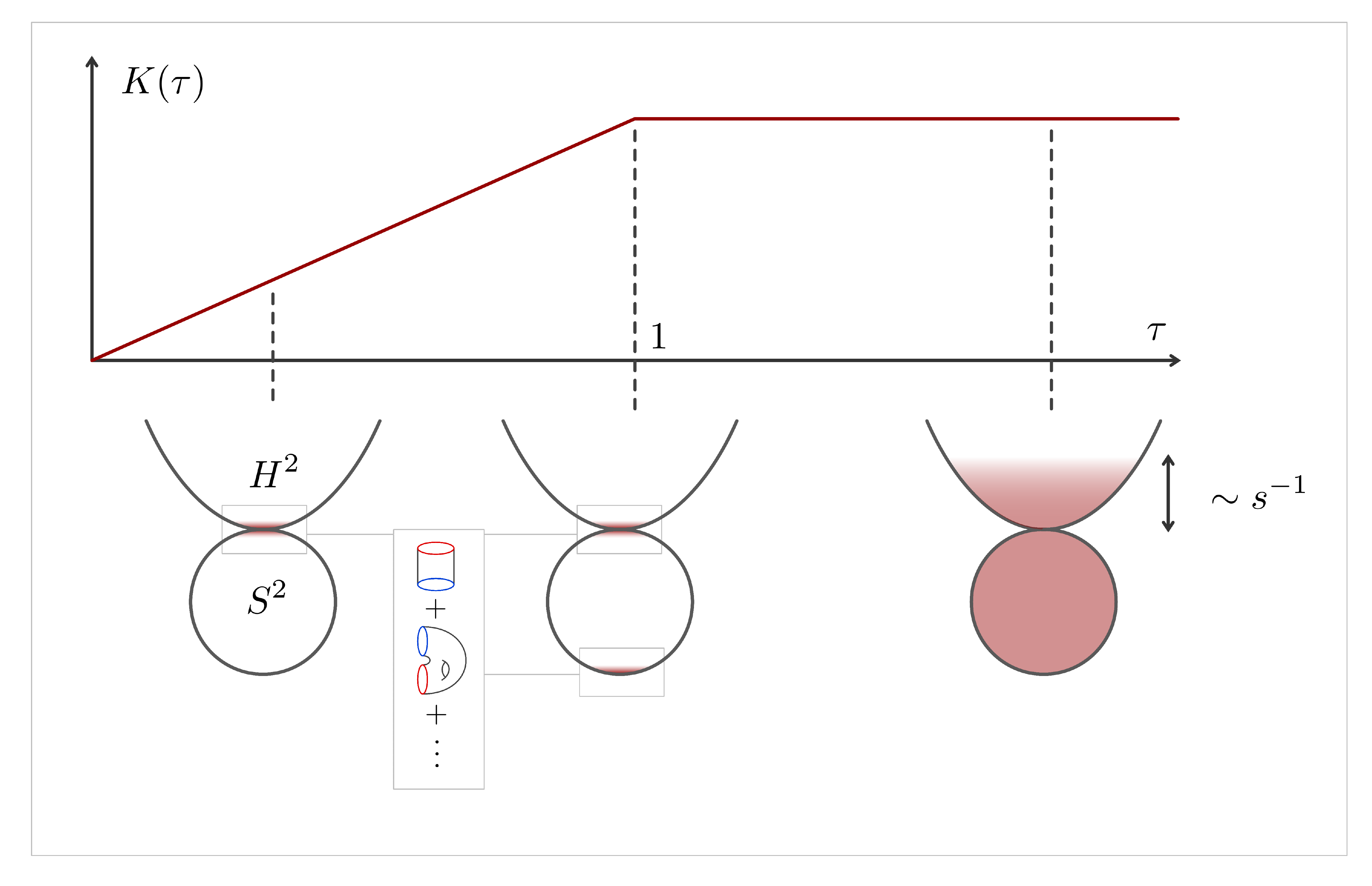}\\
\end{center}
\caption{The generic late-time behavior described by the effective theory of quantum chaos. The top line shows the ramp and plateau behavior of a simple observable -- the spectral form factor $K(\tau)$ -- and the bottom line shows how the different qualitative regimes are obtained in the sigma model. The ramp behavior is initially well described by small perturbative corrections around the standard saddle (the north pole of the sphere in the sigma model geometry $H_2\times S^2$). In this phase causal symmetry is spontaneously broken. Around the Heisenberg time $\tau = 1$, one must also include the non-perturbative contributions of the second saddle,  as described in  \eqref{eq.TheFullStructure}. At very late times the system explores the full Goldstone manifold $H_2\times S^2$ and the causal symmetry is restored. }
 \label{fig:FormFactorVsGeometry}
\end{figure}

A major advantage of the  EFT approach is that it provides access to the long time
asymptotics of spectral correlations: for $s\lesssim 1$, corresponding to energies
$\omega< \Delta$ or times $t>t_H$, the above perturbative expansion breaks down.
Instead,  the spectral correlation functions now probe the full Goldstone mode
manifold. One may thus interpret the exploration of the full manifold in terms of the
restoration of the causal symmetry. This symmetry restoration is natural from our
earlier intuitive picture of the breaking of this symmetry: the symmetry
\emph{breaking} was reflected by the replacement of a discrete pole structure by a
continuous cut at the level of the mean-field theory. However, once we probe finer
structures $\omega\lesssim \Delta$, the theory  is capable of reproducing the fact
that the spectra of individual systems are discrete, and must therefore undo the
effects of the symmetry breaking. It does so by allowing the Goldstone modes to
explore the full coset space (see Fig.~\ref{fig:FormFactorVsGeometry} for an illustration.)

Technically, the Goldstone mode matrix integrals Eq.~\eqref{eq:QMatrixIntegral} are simple enough to be doable in closed form for all symmetry
classes~\cite{EfetovSigma}. However, for the present purposes it is not necessary to
delve into the technicalities of these computations, a stationary phase analysis extended for the presence of the Altshuler-Andreev saddle, $Q_\mathrm{AA}$, defined in the beginning of the section suffices for the present purposes. 
Referring the Reader to section~\ref{sec:RMTSigmaModel} for the technical fine print, a repetition of the perturbative expansion, now around $Q_\mathrm{AA}$ gives rise to the full structure
\begin{align}\label{eq.TheFullStructure}
\includegraphics[scale=0.35]{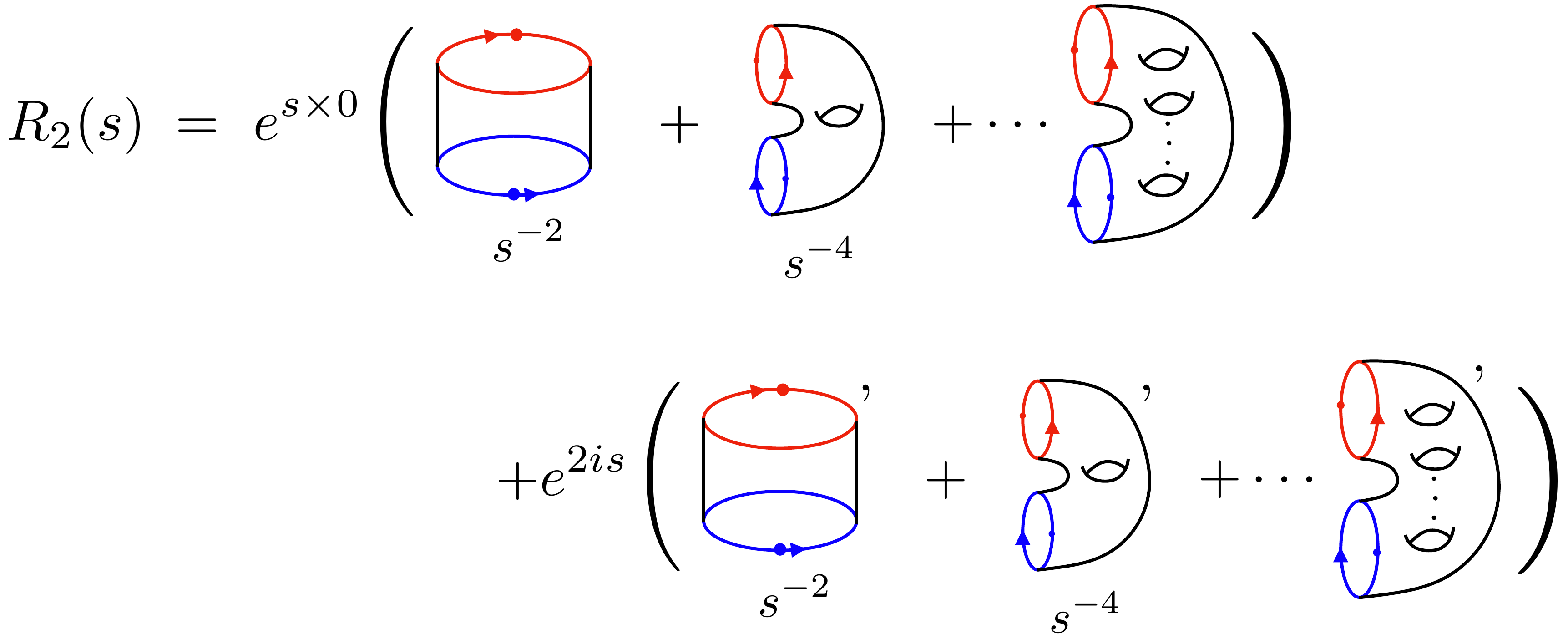},
\end{align}
where for simplicity we included orientable diagrams/surfaces only. In the semiclassically exact unitary class, the expansion truncates after the first contributions, and we obtain the spectral two point function as  
\begin{align}
    \label{eq.SpectralTwoPointUnitary}
    R_2(s)=-\mathrm{Re}\,\frac{1}{2s^2}(1-e^{-2is})=-\frac{\sin^2 s}{s^2}.
\end{align}
As the above graphical representation suggests (see also Figure \ref{fig:FormFactorVsGeometry} for a description in terms of the target space geometry of the sigma model), the perturbative expansions around
the two saddle points differ quantitatively, however they are organised in terms of
topologically equivalent surfaces, which we indicate by the primes after each
contribution around the AA saddle. In summary then, the late-time behavior of chaotic
quantum systems can understood semiclassically as an expansion around two saddles,
with the perturbative series around each saddle organised into a topological series
that can be interpreted as `wormhole' and `baby-universe' type contributions. 

 We will
elaborate on this connection to bulk physics in Section
\ref{sec.CausalSymmetryBreakingBulk} below, where we identify two-dimensional and
three-dimensional bulk spacetimes corresponding to the universal $1/s^2$ singular
diagram around the standard saddle above. It would be natural to interpret this as a
holographic version of the periodic-orbit interpretation of spectral rigidity in
quantum chaotic systems (see for example the book \cite{haake1991quantum}).

In view
of this relatively straightforward semiclassical interpretation of the expansion
around the standard saddle, $Q_0$, one may ask if the expansion around the
Altshuler-Andreev saddle, $Q_{\rm AA}$, has a similar semiclassical interpretation,
as well. This is particularly compelling in light of the fact that we can map one
saddle-point contribution to the other using the Weyl symmetry as explained. The
question is then, whether the topological expansion in $s^{-1}$ can also be
interpreted term by term in terms of something like semiclassical orbits (and by
extension, semiclassical bulk configurations). The answer is cautiously affirmative,
although that interpretation is less well established and makes use of the so-called
\emph{Riemann-Siegel lookalike}~\cite{Keating92}: First note that the semiclassical
expansion in $s^{-1}$ is an asymptotic one. Its divergence for small $s$ reflects the
exponentially growing number of long 
loops contributing to the trace of the resolvent, $W(z)$. On the other hand,
the sum is the semiclassical approximation to something finite, the density of
states, or the determinant of a random operator. Inspection of
Eq.~\eqref{eq.SpectralTwoPointUnitary}, and in particular of the negative sign
multiplying the contribution from the AA saddle suggest an interpretation of orbits
`subtracted' from the full  contents of these quantities (whatever that might be).
These vague remarks can be made much more rigorous for certain proxies of chaotic
systems such as $L$ dimensional random unitary matrices, $U$~\cite{Braun_2012}. In
this case, the role of the spectral determinants is taken by $\det(1-z U)$. These
determinants are $L$-dimensional polynomials in $z$, where the secular coefficients,
$A_n$, afford a representation as polynomials in $\tr(U^l)$, $l\le n,$ (the unitary
analog of closed loops of length $l$.) Now, unitarity requires $A_n = \bar A_{N-n}
A_N$. This key formula states that the contents of short orbits $0<n\le N/2$
determines that of the longest orbits, $N/2\le n\le N$. Using this formula, and the
aforementioned Weyl symmetry of the spectral determinant, the full spectral form
factor of unitary random maps becomes accessible via perturbative expansion. Similar
unitarity principles should apply to more complex chaotic systems and determine their
non-perturbative spectral correlations. However, the mathematically sound
implementation of these principles remains to be spelled out in practise.

\section{EFT for matrix models}
\label{sec:RMTSigmaModel}
In this section, we discuss the EFT approach in the context of matrix models of
arbitrary potential. Matrix models provide a valuable class of examples where the
effective field theory of late-time chaos can be derived from first principles,
serving as an explicit arena in which to illustrate each of the steps in Section
\ref{sec.EFTofChaos} in full detail. At the same time, they represent  duals to bulk
theories\cite{Ginsparg:1993is,Seiberg:2004at,Saad:2019lba}, indicating that the
symmetry breaking mechanism central to the present approach, too, has manifestations
in the bulk. Before addressing this perspective in
section~\ref{sec.CausalSymmetryBreakingBulk},    the two main goals of this section
are 1) to explain in technical detail some of the steps in treating the $Q$-matrix
theory and 2) to establish a set of examples in which the EFT of quantum chaos can be
derived explicitly in order to gain some more intuition about its structure. In doing
so we show in generality how to go from the $L\times L$ `color-matrix' description of
the model to the graded flavor representation, where the rank of the matrix is
${\cal O}(1)$. (As we are solely mapping a high dimensional matrix integral to a
single low dimensional one, the terminology `field theory' may be a misnomer in the
present setting. However, in view of the fact that the  flavor integrals discussed
here approximate higher dimensional field theories in the ergodic limit we keep using
it.)

\subsection{Invariant matrix models}
 An (invariant) matrix model is defined by a an ensemble of $L\times L$ random matrix Hamiltonians $H=\{H_{\mu\nu}\}$ governed by a probability distribution $P(H)dH=\exp(-V(H))dH$, where $V(H)=V(U^\dagger H U)$ is a unitarily invariant scalar function, and $dH$ a flat measure (possibly constrained to subsets of the hermitian matrices in cases where $H$ carries symmetries besides hermiticity). Assuming that $V(H)=\tr(F(H))$ is expressed via the trace of a matrix function,  we note the additional symmetry $V(H)=V(H^T)$ which we will use later in our construction.

Once again we focus  on the spectral determinant \eqref{eq.SpectralDets}, now averaged over the invariant distribution. We will demonstrate that this quantity can be exactly rewritten as a reduced integral over
graded matrices, $A$, of much lower dimension $4=2\times 2$ . Here, one factor of two
accounts for the different causality $\pm$ of the Green functions, and the second
reflects the $\mathbb{Z}_2$ grading, i.e. the presence of commuting and anticommuting variables in $A$
required to generate determinants in the denominator and numerator, respectively. The $A$-matrix integral is form equivalent
to the $H$-integral in that the integration is over the same distribution $P(A)dA=\exp(- V(A)) dA$ and a structurally identical integrand. However, it is much easier to handle, especially when it comes to the description of
correlations at `microscopic' scales of the order of the level spacing.

\subsection{Construction of the theory}
As explained in section \ref{sec.EFTofChaos} above, our starting point is the Gaussian integral representation, Eq. \eqref{eq.GaussianGradedGenFun}, of the spectral determinant Eq.~\eqref{eq.SpectralDets}, now integrated over the invariant ensemble
\begin{align}
    \label{eq:SpectralDeterminantGaussian}
  \left\langle  \mathcal{Z}(\hat z) \right\rangle= \int dH \int d(\bar\psi,\psi)\, e^{-V(H)+i\bar \psi(\hat  z-H)\psi}\,.
\end{align}
Let us write $\psi=(\phi^+,\phi^-,\eta^+,\eta^-)^T$, $\bar\psi=(\bar\phi^+,-\bar \phi^-,\bar \eta^+,\bar \eta^-)$,  so that $\phi^\pm$ denote c-number integration variables, $\bar \phi^-$ their conjugates (note the minus sign in front of $\bar \phi^-$ required by convergence),  $\eta^\pm,\bar\eta^\pm$ are independent Grassmann variables, and we have made the $H$-average explicit. We arrange the energies as a matrix,
\begin{align}
    \label{eq:MatrixIntegralHPsiRepresentation}
    \hat z=E+\hat \omega+i\delta \tau_3, \qquad \hat \omega=\text{diag}(\omega_1,\omega_2,\omega_3,\omega_4),
\end{align}
where the energy arguments have been split into $E \mathds{1}_{4\times 4}$, and the matrix of energy differences $\omega$, and we  assume $\sum_i \omega_i=0$ without loss of generality. The integral above is set up such that integration over the commuting (anti-commuting) variables produces the determinants in the denominator (numerator) entering the spectral determinant. Note that the integration vectors $\psi=\{\psi_\mu^a\}$ and $\bar \psi=\{\bar \psi_\mu^a\}$
live in a tensor product space, $\mathcal{H}_\mathrm{C}\otimes
\mathcal{H}_\mathrm{F}$, where $\mathcal{H}_\mathrm{C}$ is the $L$-dimensional
`color' representation space of the Hamiltonian and $\mathcal{H}_\mathrm{F}$ the four
dimensional graded `flavor' space of $\psi$'s internal degrees of freedom. Defined in
this way,  $\psi$ carries two natural group representations. The first is a unitary
representation, 
\be\label{eq.UnitaryColorMatrix}
\psi\to U \psi, \qquad  \textrm{where}\qquad U=\{U_{\mu\nu}\}\in \mathrm{U}(L)
\ee
Due to the invariance of the distribution, $P(H) dH =
P(UHU^{-1})dH$, this defines an exact symmetry of the integral \eqref{eq:SpectralDeterminantGaussian}. The second is a
representation under graded flavor matrices,
 \be
 \psi\to T \psi\,, \bar \psi \to \bar
\psi T^{-1}\,, \qquad T=\{T^{ab}\}\in \mathrm{GL}(2|2)\,.
\ee
 Invariance under this transformation is weakly broken by differences between the frequency arguments,
$\omega_i$, and infinitesimally by the causal increment, $i\delta$. We will see that
these two symmetries and their fate under $H$-averaging essentially determine the matrix integral.

\subsubsection{The color flavor map}
To prepare the average over $P(H)$, we define the two matrix structures, 
\begin{align}
    \label{eq:XiADef}
    \Xi\equiv  \bar \psi \psi=\{\bar\psi_\mu^a \psi_\nu^a\},\qquad \Pi\equiv \psi \bar \psi =\{\psi^a_\mu \bar \psi^b_\mu\},
\end{align}
where the first is an $L\times L$ color matrix transforming as a  flavor singlet, and the second a $4\times 4$  graded flavor matrix transforming as a color singlet.  The $H$-dependent term in the integrand can be written in the form $\bar \psi H \psi =\tr_\co (H^T \Xi)$, and the average over $H$ yields
\begin{align}
    \left\langle e^{i \bar \psi H \psi }\right\rangle_H =\left\langle e^{i\, \tr(H^T \Xi)}\right\rangle_H=\left\langle e^{i\, \tr(H \Xi)}\right\rangle_H\equiv G(\Xi),
\end{align}
where in the third equality we used the invariance of the distribution under $H\to H^T$, and in the final one defined $G(\Xi)$ as the generating function of the distribution $P(H)$. For the purposes of the present construction, there is no need to know the function $G$ explicitly. However, the above mentioned unitary invariance \eqref{eq.UnitaryColorMatrix} implies $G(U^{-1} \Xi U)=G(\Xi)$. In symbolic notation, we assume this invariance condition to be realized as $G(\Xi)=G([\tr_\co(\Xi^n)]^m)$, i.e. via dependence of $G$ on arbitrary powers of traces of powers of $\Xi$.
The key relation now coming into play is
\begin{align}
    \label{eq:TraceDuality}
    \tr_\co(\Xi^n)=\str_\fl(\Pi^n).
\end{align}
The identity is proven by the cyclic exchanges of the $\psi^a_\mu$ fields in writing out the color trace explicitly.

As a consequence of this relation, we have $G(\Xi)=G(\Pi)$, where in a slight abuse of notation we denote the function $G(\Pi)=G([\str_\fl(\Pi^n)]^m)$ by the same symbol, $G$. We may now pass back from the generating function to a distribution as
\begin{align*}
     G(\Delta)\equiv \left\langle e^{i\,\str_\fl(A\Pi)}\right\rangle_A
 \end{align*} 
 where we introduce the average
 \be
 \langle \dots\rangle_A=\int dA\, P(A)(\dots)\,,
 \ee
 over the four-dimensional graded matrix, $A$ with respect to the flat measure $dA$,
 meaning an independent integration over all commuting and anticommuting variables.
 Here, $P(A)$ is the flavor matrix distribution defined by the generating function
 $G(\Pi)$. Since $G$ did not change its form in passing from $\Xi$ to $\Pi$, the same
 it true for the distribution $P$ in passing from $H$ to $A$. Note that for the simple example of a Gaussian matrix ensemble,
\begin{align}
    \label{eq.GaussianPotential}
    V(H)=\frac{L}{g^2}\tr(H^2),
\end{align}
the
 equivalence
 
can be verified by elementary manipulation of Gaussian integrals. Substituting the result back into the starting expression, we obtain 
\begin{align}
\label{eq:MatrixIntegralAPsiRepresentation}
  \left\langle  \mathcal{Z}(\hat z) \right\rangle= \int dA \int d(\bar\psi,\psi) e^{-V(A)+i\bar \psi( z-A)\psi},
\end{align}
an expression identical to Eq.~\eqref{eq:MatrixIntegralHPsiRepresentation}, except that the integral  over the $L\times L$ Hamiltonian $H$ is replaced by one over the much smaller flavor matrix $A$. The advantage of these manipulations become obvious once we integrate over $\psi$ 
\begin{align}
  \left\langle  \mathcal{Z}(\hat z) \right\rangle= \left\langle  \mathrm{sdet}(\hat z-A)^L\right\rangle_A= \left\langle  e^{-L\,\str \ln(\hat z-A)}\right\rangle_A,
\end{align}
to obtain a (super-)determinant\footnote{The super-determinant of a graded matrices is defined as $\mathrm{sdet}\left(\begin{smallmatrix}
    a&\rho \cr \tau &b&
\end{smallmatrix}  \right)=\det a/\det(d-\tau a^{-1}\rho)$.} raised to the $L$-th
 power due to fact that  the integrand is a singlet with respect to its Hilbert-space
 indices. Making the averaging procedure explicit, we obtain a dual representation of
 the integral
\begin{align}
    \label{eq:AFunctional}
   \left\langle    \mathcal{Z}(\hat z) \right\rangle=\int dA\,e^{-V(A)-L\str \ln(z-A)},
\end{align}
now formulated in terms of the low dimensional flavor matrices.
From now on, most traces will be over flavor space, and we omit the corresponding subscript. 
 
 Building on this representation, we may now retrace  the individual steps outlined
 in section \ref{sec.EFTofChaos}, notably the  GL$(2|2)$ symmetry, and its
 spontaneous broking down to GL$(1|1)\times$GL$(1|1)$. We will go quickly through the
 derivation of the EFT building on this symmetry breaking scenario, but emphasize
 certain technical details that were left out in the more conceptual treatment above.

\subsubsection{Deriving the effective sigma model}\label{sec.DerivingEffectiveSigmaModel}
The presence of the large pre-factor $L$ in the exponent motivates a stationary phase analysis of the integral. Variation of the action yields the saddle point equation
\begin{align}
    \delta_A S(\bar A)=V'(\bar A)- L \frac{1}{\hat z-\bar A}=0 .
\end{align}
To get the spectral density, we differentiate the spectral determinant once in energy to obtain,
\begin{eqnarray}
    \label{eq:DoSSourceDifferentiation}
     -\partial_{\omega_1}\big|_{\omega=0}  \left\langle \mathcal{Z}(\hat z) \right\rangle &=& \left\langle \tr_\co\left(\frac{1}{E^+-H}\right)\right\rangle_H \nonumber \\
      &\simeq & \left(\frac{N}{E+i\delta \tau_3-\bar A}\right)_{11},
 \end{eqnarray} 
 where in the second line we have evaluated the expression at the stationary point. The approximate equality sign indicates that the relations holds up to $1/L$ corrections. Accordingly, the average resolvent  is straightforwardly obtained by solving the
 stationary phase equation at $\omega=0$. Specifically, the average spectral density, $\rho(E)$, follows from taking the imaginary part of \eqref{eq:DoSSourceDifferentiation} evaluated at the saddle point. To illustrate this point, consider the case of the Gaussian matrix
 potential, $V(A) = \frac{L}{g^2}{\rm Tr} (A^2)$, for which the variational equation
 at $\omega=0$ takes the form
\begin{align*}  
  g^2 \bar A-(E+i\delta \tau_3 -\bar A)^{-1}=0\,.
 \end{align*} 
Reflecting the rotational symmetry of the action,  this equation affords solution in terms of diagonal matrices $\bar A$. It is a  quadratic equation, individually for each of the diagonal matrix elements, and we need to pick one of two solutions --- this is where the spontaneous symmetry breaking happens. Specifically, for $|E|<2g$, we have the solutions $\bar A(E)=E/2\pm i\gamma (E)$, and the `natural' of these is 
\begin{align}   \label{eq.AGaussianSolution}
  \bar A(E)=\frac{E}{2}+i
 \gamma(E)\tau_3, \qquad \textrm{where} \qquad \gamma(E)=\sqrt{g^2-
 \left(\frac{E}{2}\right)^2 }.
\end{align} 
Physically, this solution is natural in that the sign of the imaginary part $i\gamma$ is dictated by the infinitesimal $i \delta$. We interpret this as shifting of a pole into the complex plane reflecting the smearing of a discrete pole structure into a cut at mean field level\footnote{In the bosonic sector of indices diagonal indices $|a|=1$, this sign choice actually is obligatory. The reason is that the saddle points with their finite imaginary part must be reached by deformation of the real integration contours defined by the eigenvalues of the Hermitian integration variables $A^{ab}$, $|a|=|b|=1$. Under the logarithm, these variables appear infinitesimally shifted into the upper/lower complex half plane, depending on the sign $\pm i\delta$. Reaching the `wrong' saddle points would require passage through the cut of the logarithm and cause a a divergence in the integral. (This is best seen in the representation, $\exp (- \ln(E\pm i \delta + x))=1/(E\pm i  \delta + x)$.) In the fermionic sector, no such problem exists, and either saddle point is reachable. }.  Substitution of this variational
 solution into Eq.~\eqref{eq.Z2Intro} leads to the famous semicircular
 density of states
 \begin{align}   \label{eq.SemicircularDOS}
     \rho(E)=-\frac{1}{\pi}\mathrm{Im}\,\tr_\co(G^+(E))=\frac{L}{\pi g}\left(1-\left(\frac{E}{2g}\right)^2\right)^{1/2}.
 \end{align}
Returning  to the case of an arbitrary invariant potential, suppose  that a solution to the saddle-point equation of the form
\begin{align}
    \label{eq:SaddlePointSolution}
    \bar A(E)\equiv \epsilon(E)+i\gamma(E)\tau_3
\end{align}
 with real $\epsilon,\gamma$ has been found. We then build on the spontaneous breaking of the causal symmetry breaking reflected by the $\pm \gamma(E)$ term, and turn to step 3 in constructing the EFT which instructs us to define the matrix $Q$
 \begin{align}
    \label{eq:StationaryAGeneralized}
    \bar A\to  T \bar A T^{-1}\equiv \epsilon(E)+i \gamma(E) Q, \qquad Q\equiv T \tau_3 T^{-1}\,,
\end{align}
 parametrizing the Goldstone manifold \eqref{eq.GoldstoneCoset} of stationary solutions, where $T$ has been defined in \eqref{eq.ExponentiatedPions}. Moving along the general program we can now write down the effective action using this object.

\vskip1em
\noindent \textit{Symmetry breaking and effective action:}  we begin with the  substitution of Eq.~\eqref{eq:StationaryAGeneralized}  into the action of
Eq.~\eqref{eq:AFunctional}. The $\mathrm{GL}(2|2)$ invariance of the potential term,
$V(A)=V(TAT^{-1})$, implies the decoupling of the former from the Goldstone mode
integral. (This is an elegant way of seeing why the singular Goldstone mode fluctuations are oblivious of the detailed form of the invariant distribution. Their job solely is to determine the average spectral density via the above $\gamma(E)$.) The action  thus reduces to
\begin{align*}
  S[Q]=L \,\str\ln\left(\tilde
 E+\hat z+i\gamma   Q\right),
 \end{align*}
  where  we redefined $\hat z\to \hat z-E$ to single out the symmetry breaking parameters and absorbed the real part of the
stationary point solution into a redefined energy parameter $\tilde E=E-\epsilon(E)$. Using
the  cyclic invariance of the trace, $S[Q]=L \str\ln\left(\tilde E+T^{-1}\hat z T+i
 \gamma \tau_3\right)$,  we couple the Goldstone mode fluctuations to the
explicit symmetry breaking, $z$. In a final step,  we expand to first order in
$z/\gamma$ and use the saddle point property $(\tilde E+i \gamma
\tau_3)^{-1}=(E-\bar A)^{-1} = \frac{i \pi}{L} \rho \tau_3+\dots$, where $(\dots)$ are contributions proportional to the unit matrix to obtain the effective action
\begin{align}
    \label{SigmaModelAction}
    S[Q]=i\pi \rho\,\str(\hat z Q).
\end{align}
It should come as no surprise that this coincides with the ergodic limit
\eqref{eq.SErgodic} of the EFT for extended systems.  We are now in a position to
explore the physics of the Goldstone modes in the setting of a model which
does not contain any `spatially fluctuating' modes. This goes over the same physics
as as visualized in \ref{sec.EFTofChaos} using ribbon graphs, but we wish to take the
opportunity here to give a careful derivation of all the prefactors and signs that we
had previously glossed over.

\subsection{The integral over the Goldstone manifold}
In this section, we turn to the details of the $Q$-matrix integration over the identical actions Eq.~\eqref{SigmaModelAction}, or Eq.~\eqref{eq.SErgodic}. Their equality implies that the results derived here equally apply to the matrix model and to the physics of extended systems below their Thouless energy. 
\subsubsection{One-point functions: the spectral density}
In general, the method of choice  for doing the $Q$ integration depends on both, the
magnitude of $\omega$, and the specific observable to be computed, reflected in a
judicious choice of the sources for advanced and or retarded correlation functions.
Note that in our previous treatment the source structure was reflected in the way the
projectors $P^\pm$ appeared in correlators such as
\eqref{eq.GoldstoneCylinder}.  In the following, we address three instances of
interesting  source insertions and $\omega$-ranges, and along the way compute the
ramp-plateau profile of GUE spectral statistics.

The evaluation of the integral is particularly easy in cases where the integrand
possesses a fully unbroken supersymmetry, i.e. invariance under a supersubgroup
$\mathrm{GL}(1|1)$. Under these circumstances, a theorem due to Efetov and Wegner  states that the integral
collapses to its value at the `coset origin',\footnote{The rational behind this
collapse is that in the case of a supersymmetry, we have a conflict of interests: due
to the invariance, the integrand does not depend on the Grassmann valued generators,
$\eta$, of the symmetry, and the integration $\int d\eta=0$ suggests a vanishing
integral. On the other hand, it also does not depend on the non-compact bosonic, $x$,
variable of the symmetry subgroup, and the integral $\int dx=\infty$ suggests a
divergence. The solomonic resolution of the $0\times \infty$ conflict is a collapse
of the integral to configurations where all generators vanish. For a pedestrian
discussion of the details, we refer to Ref. \cite{haake1991quantum}.}
\begin{align}
    \int dQ e^{-S[Q]}\stackrel{\mathrm{susy}}{=}e^{-S[\tau_3]}.
\end{align}
As an example, consider the case of absent sources, and degenerate energy arguments $\hat z= i\delta \tau_3$. In this case, the spectral determinants cancel out, and $\langle\mathcal{Z}(\hat z)\rangle=1$. This is confirmed as
\begin{align}
    \left\langle{\cal Z}_{(4)}(0,0,0,0)\right\rangle=e^{- \delta\pi \rho\,\str(\tau_3)}=1,
\end{align}
where the action reads more explicitly $\str(\tau_3):=\str(\tau_3 \otimes \id)=1$, by supersymmetry. Differentiating once with respect to sources, as in \eqref{eq:DoSSourceDifferentiation}, confirms the identification of the prefactor of the action as the density of states $\rho$, which can be evaluated as demonstrated in \ref{sec.DerivingEffectiveSigmaModel} above.
\subsubsection{Perturbative spectral correlations: the ramp}\label{sec.SolvingTheRamp}
We next apply the formalism to the computation of \textit{spectral fluctuations}, as described by the cumulative spectral two-point function \eqref{eq.Z4Intro}. We rewrite \eqref{eq.Z4Intro} slightly, using the definition of the mean level spacing as
\begin{eqnarray}
    R_2(\epsilon) &=&  \Delta^2 \left\langle\rho\left(E+\frac{\epsilon}{2}\right)\rho\left(E-\frac{\epsilon}{2}\right)\right\rangle_c\nonumber\\
    &=&\frac{\Delta^2}{2\pi^2}\mathrm{Re}\left\langle\mathrm{Tr}(G^+(E+\tfrac{\epsilon}{2})\,\mathrm{Tr}(G^-(E-\tfrac{\epsilon}{2})\right\rangle_c,
\end{eqnarray}
where we used that the connected average $\langle G^\pm G^\pm\rangle_c\equiv \langle G^\pm G^\pm\rangle-\langle G^\pm\rangle\langle G^\pm\rangle$ of Green functions of coinciding causality vanishes. Inspection of the spectral determinants shows that the average of Green functions in this expression is obtained as
\begin{align}
    R_2(\epsilon)=\frac{\Delta^2}{2\pi^2}\mathrm{Re}\,\partial^2_{\omega^{\mathrm{f}+},\omega^{\mathrm{f}-}}  \left\langle \mathcal{Z}(\hat z)\right\rangle_c,
\end{align}
where the derivative is evaluated at the configuration $\omega = \tfrac{\epsilon}{2}\tau_3$, and the subtraction of a `disconnected' contribution to the functional integral, indicated by the subscript $c$, implements the cumulative average. Doing the derivative in the representation~\eqref{eq:QMatrixIntegral}, we arrive at the representation 
\begin{eqnarray}
    \label{R2QMatrixIntegral}
    R_2(\epsilon)&=& -\frac{1}{2}\mathrm{Re}\left\langle \str(Q(P^+\otimes P^\mathrm{f}))\,\str(Q( P^{-}\otimes P^\mathrm{f}))\right\rangle_{Q,c},\nonumber\\
     &&\langle\dots\rangle_Q\equiv \int dQ \,e^{-i \frac{s}{2}\,\str(Q\tau_3)}(\dots),
\end{eqnarray}
where $P^s, (s=\pm)$ projects on the advanced and retarded causal sector,
 respectively, while $P^{\mathrm{f,b}}$ project on the fermionic and bosonic sector,
 respectively, and we represented the action (after source differentiation) in the dimensionless units Eq.~\eqref{eq.sTauDef}.
 This is the matrix theory whose topological expansion, ordered by powers
 of $s$, we discussed in Section \ref{sec.EFTofChaos}. We now carry out the integral
 explicitly, thus   fixing the  coefficients of each term in that
 expansion. 
 In a first step, we concentrate on large energy offsets, $\epsilon\gg
 \Delta$ or $s\gg 1$. In this case, the action of the matrix integral is strongly
 oscillatory, and Goldstone mode fluctuations are confined to small neighborhoods $\sim s^{-1}$ of the stationary points on the saddle point manifold. These residual fluctuations
 are conveniently described in an  (exponential) parameterization,
\begin{align}
    Q=T\tau_3 T^{-1},\qquad T=\exp W,\qquad W=
    \left(\begin{matrix} &B\\  \tilde B&\end{matrix} \right),
\end{align}
where the block structure implements the anti-commutativity $[W,\tau_3]_+$ of the  generators with the coset origin, and the $2\times 2$ super-matrices
\begin{align}
    B=\left(\begin{matrix} z&\mu \\ \nu&w 
     \end{matrix}\right),\qquad
     \tilde B=\left(\begin{matrix} \bar z&\bar\nu\\ \bar\mu&-\bar w   \end{matrix}\right),
\end{align}
contain the two complex commuting ($z,w$) and four Grassmann ($\mu,\dots,\bar\nu$) integration variables of the model, parametrizing the most general Goldstone fluctuation. As we stated earlier in more general terms, we can see explicitly that in the bosonic sector, $B^\mathrm{bb}$ the variable $z$ spans a hyperboloid with radial coordinate $|z|$ and in the fermionic sector $B^\mathrm{ff}$ the variable $w$ a sphere, $Q^{\mathrm{ff}}=n^i \tau_i$, with $|n|=1$, where $|w|=\pi/2$ represents a rotation from the north pole, $n=(0,0,1)^T$, to the south pole, $n=(0,0,-1)^T$, i.e. the transformation $T_W$ mapping the standard onto the AA saddle. Let us now substitute an expansion to quadratic order in fluctuations
\begin{align*}
    Q=\tau_3(1-2W + 2W^2+\dots),
\end{align*}
 into Eq.~\eqref{R2QMatrixIntegral}, we obtain up to quadratic order the expression,
\begin{align*}
    R_2(\epsilon)=-2\,\mathrm{Re} \left\langle \str(B\tilde B P^\mathrm{f})\,\str(\tilde B BP^\mathrm{f}
)\right\rangle,
\end{align*}
where the $B$-average is defined in Eq.~\eqref{eq.BAverage}, and  the disconnected
pure saddle point contribution, $B=0$, does not contribute to the cumulative average. Note that we have now recovered in detail the structure we had described in a more qualitative setting in Eqs. \eqref{eq.GoldstoneCylinder} -- \eqref{eq.TwoCrossesOrientable} above.
Doing the final Gaussian integral\footnote{This can be done by brute force, or using
the matrix version of Wick's theorem, $\langle\str(B X\tilde B
Y)\rangle=\frac{1}{is}\str(X)\str(Y) $ , $\langle\str(B X)\str(\tilde B
Y)\rangle=\frac{1}{is}\str(X Y)$.}, we obtain
\begin{align}
\label{R2Perturbative}
    R_2(s)\simeq - \frac{1}{2s^2},
\end{align}
corresponding to the ramp, $K(\tau)=\tau$, upon Fourier transformation to dimensionless time. The strategy for refining this result beyond the leading order in the parameter $s^{-1}\ll 1$ is evident: expansion of the action in the generators $B$ leads to terms $\sim s\,\str(B\tilde B)^n$, which after integration contribute as $s^{1-n}$. However, in the unitary symmetry class, it turns out that the prefactors of all these contributions cancel out order-by-order in the $s^{-1}$ expansion, and that~\eqref{R2Perturbative} does not change in perturbation theory. 

\subsubsection{Non-perturbative correlations: Weyl symmetry and the plateau}
For $s<1$, the stationary phase approach to the Goldstone mode is no longer parametrically controlled. At the same time, the absence of perturbative corrections to the approximation Eq.~\eqref{R2Perturbative} hints at a `semiclassically exact' integral. Heuristically, the underlying mechanism can be understood by inspection of the fermionic sector of the theory: with
$Q^\mathrm{ff}=n^i \tau_i$, the action $\mathrm{tr}(Q^\mathrm{ff})=2n_3$ is
proportional to the height function $n_3=\cos(\theta)$ on the sphere, and $\int
dQ^\mathrm{ff}$ an integration over the canonical measure. The integral thus affords
an interpretation as partition sum of a spin precessing in a fictitious magnetic
field of strength $s$. This partition sum is a classic example of the semiclassical
exactness principle \cite{Blau:1995rs}, and it turns out that this feature carries over to its
supersymmetric extension\footnote{ Recall that an integral is semiclassically exact
if it assumes the form of a partition sum over an effective Hamiltonian,  all whose
trajectories are periodic and have equal revolution time. Both conditions are met in
the present context: the Goldstone mode manifold is symplectic with $\omega \equiv
\str(QdQ\wedge dQ)$ as defining  two form.
The integration extends over the symplectic measure, and the integrand contains the exponentiated Hamiltonian $H\equiv-is\, \str(Q\tau_3)$. For a given initial point. $Q$, the trajectories $Q(t)$ satisfy the condition $\iota_{\dot Q}\omega= 2\,\str(Q\dot Q dQ) =dH$, i.e. the tangent to the flow is a Hamiltonian vector field. A straightforward computation shows that for our present `spin in a magnetic field of strength $s$' Hamiltonian, this condition is met by $Q(t)=e^{is t \tau_3}Qe^{-is t\tau_3 }$  'spin precessing at constant frequency $s$'. All trajectories are closed and have equal revolution time $2\pi/s$.}. 

However, the semiclassical exactness principle requires us to account for all
stationary points of the integrand. We already mentioned the stationarity condition,
$[Q,\tau_3]=0$, which besides by the standard saddle $Q=Q_0=\tau_3$ is solved by the
AA saddle $Q=T_W Q_0 T_W^{-1}=\tau_3 \otimes \sigma_3^{\mathrm bf}$, with  stationary
action  $S(Q_{\mathrm{AA}})=i s \,\str( \tau_3\otimes
\sigma_3^{\mathrm{bf}})=2is$. Referring for details of the Gaussian integration
around the AA saddle point to the original reference \cite{PhysRevLett.75.902}, we
note that it produces the same factor as in Eq.~\eqref{R2Perturbative}, but with
inverted sign. Adding the two terms, we obtain the full result
Eq.~\eqref{eq.SpectralTwoPointUnitary}. For completeness, we mention that the full
Goldstone mode integrals required to obtain the correlation functions in other
ensembles are doable in closed form by introducing coset space `polar coordinates' and using the high degree of rotational invariance of the action. For the technical details, interested readers are referred to to Ref.~\cite{EfetovSigma}.

\section{Causal symmetry breaking in the bulk }
\label{sec.CausalSymmetryBreakingBulk}
We will now describe the emergence of the sigma-model from bulk considerations. The main ingredient to develop a bulk understanding  is, once again, the determinant operator
\begin{equation}
\det (E^\pm-H) = e^{{\rm Tr}\, {\rm log} (E^\pm-H)}\,
\end{equation}
This object should be thought of as inserting a D-brane in the bulk at position $E$ (see for example \cite{Polchinski:1994fq,Ooguri:1999bv}), and we distinguish here between an advanced brane and a retarded brane, depending on the sign of the (infinitesimal) imaginary part of the energy, as indicated. The Hermitian matrix $H$ -- the Hamiltonian from the point of view of our discussion of quantum chaos above -- corresponds to the presence of an additional {\rm stack} of D-branes, one for each dimension of the many-body Hilbert space. One way to see this is by noting that the determinant operator can be obtained as the exponential of the loop operator
\begin{equation}\label{eq.LoopOperator}
  {\cal W}(E) =   {\rm Tr}\log \left( E-H \right)\,,
\end{equation}
which acts to insert a boundary in an open string wordsheet with boundary condition such that the string now ends on the `$E$-brane'.
Upon exponentiating to form the determinant, the ascending powers of the ${\rm Tr}\log E-H $ insertion correspond to open string
world sheets with an increasing number of boundaries with the right combinatorics to count all possible ways the string can end on the D-brane \cite{Polchinski:1994fq}. To summarise then, $H$ is an
$L\times L$ matrix corresponding to the presence of a (large) stack of $L$ D-branes,
while $E$ corresponds to the position of a single additional ``probe" D-brane. Given that these additional branes are added in order to construct ratios of spectral determinants as in \eqref{eq.SpectralDets}, we sometimes refer to these objects as ``spectral branes". We
will return to more concrete microscopic realizations of such objects below in the
context of minimal string theory, where we can take the $L$ branes to be of ZZ type
and the ${\cal O}(1)$ spectral branes to be of FZZT type
\cite{Seiberg:2003nm,Seiberg:2004at,Maldacena:2004sn}. The physics we want to focus
on is that of open strings stretching between these different types of branes. Let us
introduce these degrees of freedom by writing the determinant operator explicitly as
\begin{equation}
    \det (E-H) = \int d(\bar\eta,\eta)\exp \left(\bar \eta (E-H)\eta \right)\,,
\end{equation}
where $\bar\eta,\eta$ are our usual $L-$component Grassmann vectors, now identified with fermionic open-string modes stretching between a single brane parametrized by $E$, and the $L$ branes in the stack parameterised by $H$. 
A second kind of excitation is given by bosonic strings stretching between the
branes, in which case we exponentiate the {\it inverse} determinant operator as
\begin{equation}\label{eq.ghostDet}
   \frac{1}{\det (E^\pm-H)} = \int d(\bar\phi,\phi)\exp \left(\pm i\bar\phi (E' \pm i\varepsilon-H)\phi \right)\,, 
\end{equation}
where we have been careful about adding an imaginary part $(\varepsilon>0)$ to ensure
convergence of the Gaussian integral, in the same way as in Section
\ref{sec.EFTofChaos} above. Such inverse brane determinants have been considered in the
past, and have been called anti-branes or, perhaps more appropriately as ghost branes
\cite{Okuda:2006fb}. As an alternative to using ghosts, we might employ $2R$ replica branes of the `normal' FZZT
type. However, except for the unitary class, replicas are ill suited to the description of the doubly non-perturbative limit~\cite{Verbaarschot_1985} which is why we generally prefer to work with ghosts (See Appendix~\ref{sec:supersymmetry_vs_replicas} for a discussion of replicas vs. supersymmetry from the matrix theory perspective.) 

It is thus clear how to interpret a generating functional of the type ${\cal Z}(\hat
z)$ (see Eq. \eqref{eq.SpectralDets}) in the bulk. We then have again the object \eqref{eq.SpectralDets}), but
now with an explicit {\it physical realization} of the modes $\psi = (
\phi^+,\phi^-,\eta^+,
\eta^-)$ which we had previously introduced as auxiliary objects. As we have seen, the ratio of determinants defining the
generating functions ${\cal Z}$ is the starting point to extract universal late-time
chaos from symmetry considerations alone. Given our interpretation of the $\eta$ and
$\phi$ variables used to exponentiate the D-brane operator as open string modes.
These string modes have Chan-Paton factors $\psi_\mu^a$, $\mu = 1,2\ldots L$, $a =
1,2\ldots n$ which allow them to end on any of the $L$ D-branes making up the `sea'
and/or on any of the $n$ probe branes. Note that the present
approach is supersymmetric by design, with a supergroup acting upon the
Chan-Paton degrees of freedom  (see for example \cite{Okuda:2006fb} for a general description of
such objects). Both the replica and the SUSY perspective reveal that the bulk
manifestation of the Goldstone modes of the chaotic sigma model are effective bound
states of strings
\begin{align*}
    \Pi^{ab} = \sum_{\mu=1}^L \psi_\mu^a \bar\psi_\mu^b
\end{align*}
resulting from integrating out over the `sea' degrees of freedom associated with the stack of $L$ D-branes labelled by the index $a$. This projects onto singlets under this `color' group, leaving the effective strings to transform in with their $a,b$ indices in the adjoint of the `flavor' group. For the computation of spectral correlations, for example of pair correlations \eqref{eq.SpectralCorrelationFunction}, the causal symmetry breaking mechanism applies and we may effectively concentrate on the light sector of modes contained in $\Pi$, in other words the Goldstone sector $Q$. 

\begin{figure}[h]
\begin{center}
\includegraphics[width=.6\columnwidth]{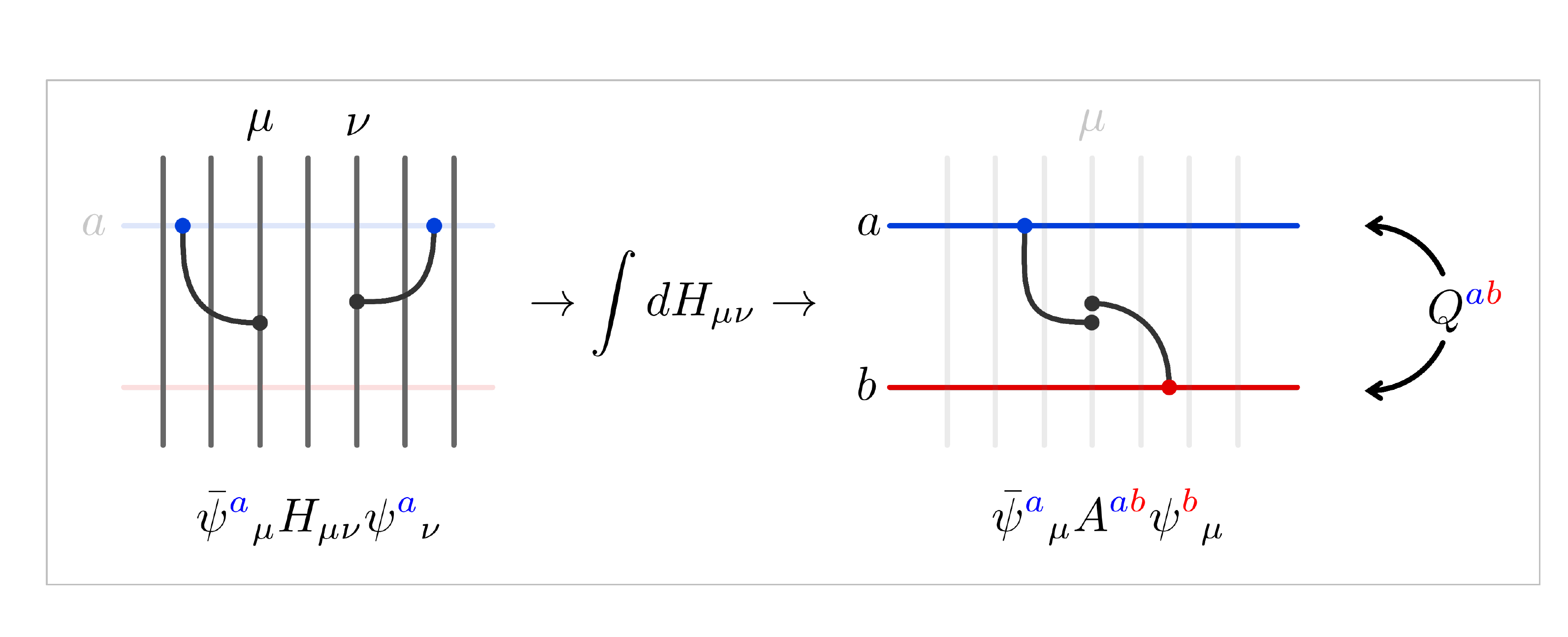}\\
\end{center}
\caption{Bulk sigma model: we start from the configuration  comprising  `sea branes', $\mu=1,\ldots L$, and spectral branes, $a=1,\ldots 4$. As a  microscopic realization, for example in the context of minimal string theory, we can take the $L$ `sea branes' to be a stack of $L$ coincident ZZ branes and the spectral branes to be FZZT branes. In the corresponding matrix theory, pairs of strings connecting sea branes $\mu$ and $\nu$, respectively, to the same spectral brane, $a$, are represented by the bilinears $\bar \psi_\mu^a H_{\mu\nu} \psi_\nu^a$, with Chan-Paton factors $\psi_\mu^a$. Integration over the stack of sea branes leads to a dual picture, with `string bound states' $\bar \psi_\mu^a A^{a,b}\psi_\mu^b$ correlating spectral branes, $a,b$, via an induced bulk geometry with each other. The previously exact symmetry between different $a$'s (for spectral branes at coinciding energy-coordinates, $E$) is spontaneously broken and leads to the emergence of the $Q^{ab}$ pseudo Goldstone degrees of freedom. The reader may note that this is a microscopic description of an open/closed duality in the presence of `flavor' branes \cite{Karch:2002sh} giving rise to an additional open string sector (the FZZT-ZZ strings).  In section \ref{sec.ContinuumGravityFormulation} we show how to extract the Goldstone contributions from a world-sheet calculation of the FZZT-ZZ strings.}
 \label{fig:BraneVsQ}
\end{figure}

This bulk interpretation of the chaotic sigma model is illustrated in Figure \ref{fig:BraneVsQ}.

\FloatBarrier
\subsection{Causal symmetry breaking in minimal string theory}\label{sec.CausalSymmetryMMST}
We now go through the exercise of describing the bulk picture of the EFT of quantum
chaos in minimal string theory. This allows us to make direct contact with the
matrix-model techniques introduced above and then compare them to the bulk 2D gravity
or `continuum worldsheet' perspective. Referring to Refs.~
\cite{Seiberg:2004at,Martinec:2004td,Mertens:2020hbs} for reviews, we note that minimal string theory is defined by coupling 2D Liouville
theory to a $(q,p)$ minimal model matter CFT (for $q,p$ two relatively prime integers). For concreteness we will be only
interested in the cases $(2,p)$, which have a dual description as one-matrix models
with invariant potentials depending on the value of $p = 2m-1\,,(m\in
\mathbb{Z})$. It turns out that minimal string theories contain D-branes of the type
we can use to implement the spectral determinant construction at the heart of our
analysis.  These D-branes can be viewed as CFT boundary states, tensoring a Liouville
boundary state, \cite{Fateev:2000ik,Seiberg:2003nm}, with a matter boundary state, or
alternatively as boundary conditions on the worldsheet theory. Reference
\cite{Mertens:2020hbs} contains an extensive review of the relevant
constructions\footnote{We also refer the interested Reader to the excellent review
\cite{Ginsparg:1993is} for more information. A brief overview from a modern
perspective is given in  \cite{Seiberg:2004at}.} from a perspective pertinent to this
work. In the following we will work both from the matrix-model perspective as well as
from the worldsheet perspective, the latter taking the role of bulk spacetime. The
former description will be very familiar from our Section \ref{sec:RMTSigmaModel},
while the latter will recover individual  universal contributions from the 2D gravity
perspective.

\subsubsection{Double scaling to the spectral edge}
One way of defining $(2,q)$ minimal string theory is via a double-scaled limit of a random matrix ensemble of a single matrix which we may think of as the Hamiltonian $H$, so that
\be
\langle {\cal Z}(\hat z) \rangle = \int dH e^{-V(H)}{\cal Z}(\hat z)\,,
\ee
falling squarely into the class of models \eqref{eq:SpectralDeterminantGaussian}
studied in Section \ref{sec:RMTSigmaModel}. However, the present discussion
requires an extra twist, we need to zoom into the vicinity of the spectral edge. To
understand what this means in the present context, first consider the generating
functional $\mathcal{Z}_{(2)}$ Eq.~\eqref{eq.Z2Intro} of the spectral density with
derivative taken at $z=E\pm i0$. Depending on the choice of $V(H)$, there will be a
subset of energies $E$ with finite spectral density --- technically, the range of
energies with symmetry broken stationary field solutions. Symbolically denoting the
width of this spectral support by $a$, and its minimum energy by $E_0$, we consider
the `double scaling limit', of a large number of levels, $L\to \infty$, for
separations off the band edge $(E-E_0)/a\to 0$. In the following we consider the
simplest case, namely  $(2,1)$ minimal string theory, whose invariant matrix
potential is Gaussian, Eq.~\eqref{eq.GaussianPotential}, to demonstrate how
this limit leads to a variant of the Kontsevich matrix model
\cite{Kontsevich:1992ti}. However, the same method is applicable to other potentials describing theories in the 
$(2,q)$ family.   For the discussion of the equivalent continuum worldsheet
approach we refer to section~\ref{sec.ContinuumGravityFormulation} below.

After integrating out the variables $\psi,\bar\psi$ (now interpreted as open string degrees of freedom) we obtain the action \eqref{eq:AFunctional}, which we repeat here for convenience:
\begin{align}\label{eq.MinimalFlavorAction}
\langle {\cal Z}(\hat z) = \int dA\, e^{-V(A) - L \,\str  \ln(\hat z+A)}\,,
\end{align}
with $V(A)=\frac{L}{g^2}\str(A^2)$, for the $(2,1)$ variant. The solution $\bar A$ of
the stationary equations is given by Eq.~\eqref{eq.AGaussianSolution} with associated
spectral density Eq.~\eqref{eq.SemicircularDOS}, indicating that $E_0=-\sqrt{2}g$
defines the edge of a spectrum of width $a=2\sqrt{2}g$. We now consider energies
close to the band edge, $z=-E_0+\zeta$. For fixed $\zeta$, we
scale $L\to \infty$ along with $g\to \infty$ such that there is still a
macroscopically large number of levels in the interval $[-E_0,-E_0+\zeta]$. In this
limit, Eq.~\eqref{eq:SaddlePointSolution} reduces to
\begin{align}
    \label{eq.SaddlePointNearEdge}
    \bar A(\zeta)=\frac{-\sqrt{2}g+\zeta}{2}+i \tau_3\frac{g^{1/2}}{2^{1/4}}\sqrt{\zeta}
\end{align}
with associated mean field spectral density
\begin{align}\label{eq.SpectralDensityEdge}
\rho(\zeta) = \frac{e^{S_0}}{\pi}\sqrt{\zeta}\,, \qquad e^{S_0}\equiv \frac{2^{3/4}L}{g^{3/2}}.
\end{align}
We obtain a two-dimensional flavor matrix model describing the fine structure of edge of single level spacings by expansion around the symmetry unbroken solution right at the edge, $\bar A(0)=\frac{g}{\sqrt{2}}$. 
 Defining $A=\bar
A(0)+\frac{\sqrt{g}}{2^{1/4}}a$ where  the scaling factor
upfront the fluctuation matrix $a$ is introduced for convenience, and expanding to leading order in $a$ and $\zeta$, we obtain
\begin{align}\label{eq.SuperKontsevich}
\langle {\cal Z}(\hat z)\rangle \simeq \int da\, e^{-e^{S_0}{\rm STr} \left(\frac{a^3}{3} + \hat \zeta a \right)}\,,
\end{align}
which is a variant of the Kontsevich model~\cite{Kontsevich:1992ti}. Equivalent
derivations of this model from double-scaled matrix theories have appeared before in
\cite{Gaiotto:2003yb,Maldacena:2004sn}, although not in terms of a supersymmetric
formalism as we have utilized here. In the absence of sources, $\hat\zeta = \zeta
\mathds{1}$,  the action is invariant under graded general linear transformations $ a
\rightarrow T a T^{-1}\,,\, T \in {\rm GL}(2|2)$, and the Efetov-Wegner theorem secures the normalization of
the partition sum $\mathcal{Z}(\zeta)=1$. 

In the presence of sources, a parameterization of $a=\left(\begin{smallmatrix}
    \lambda_1& \rho\cr \sigma &i\lambda_2
\end{smallmatrix}  \right)$ followed by integration over the Grassmann variables $\rho,\sigma$ leads to 
\begin{align}\label{eq.AiryIntegralforDOS}
    {\cal Z}(\hat \zeta) = \left( \frac{\partial}{\partial \zeta_1} - \frac{\partial}{\partial \zeta_2} \right)\int \frac{d\lambda_1 d \lambda_2}{2\pi} e^{-e^{S_0} {\rm STr}  \left(\frac{\lambda^3}{3} + \hat\zeta \lambda \right)}\,,
\end{align}
where the diagonal matrix $\lambda\equiv a|_{\rho=\sigma=0}$ contains the commuting variables, $\lambda_1,\lambda_2$ and a choice of integration contours safeguarding convergence is implicit to the definition of the integral.  Finally the factor $(2\pi)^{-1}$ ensures the proper normalization of the bosonic measure, which we did not explicitly specify in Equation \eqref{eq.ghostDet} above. In order to obtain the spectral density Eq.~\eqref{eq.Z2Intro} with $z\to \zeta$, 
a further $\zeta_2$ derivative has to be taken, before setting the energy arguments equal to each other. This results in the Airy density of sates
\bea
\label{eq.AiryDos}
    \rho(\zeta) &=& e^{2S_0/3} \left(-{\rm Ai}(x)^2 + x {\rm Ai}'(x)^2\right)\,,\qquad \left( x = -  e^{2S_0/3}\zeta\right)\nonumber\\
    & \sim & \frac{e^{S_0}}{\pi} \sqrt{\zeta}\,,
\eea
where in the second line we have used the standard asymptotics of the Airy function (see Appendix \ref{app.Airy}) to derive the leading behavior for $e^{2S_0/3}\zeta \gg 1$. It is then clear that the saddle-point evaluation we performed above is precisely the semi-classical limit $e^{S_0}\gg 1$ of the exact expression corresponding to the well-known asymptotics of the Airy function.

In principle, we may upgrade the above integral to one over $\mathrm{GL}(2|2)$
matrices to obtain the correlations in the Airy DoS right at the edge. However, for
the present purposes, it is sufficient to move somewhat into the double scaled
spectrum and expand around the symmetry broken stationary points
Eq.~\eqref{eq.SaddlePointNearEdge} at finite $e^{-S_0}\ll  \zeta \ll g$. In this
case, there is nothing left to be done, and we can cut and paste from
section~\eqref{sec:RMTSigmaModel}, up to and including the final result, the sine-kernel level correction
Eq.~\eqref{eq.SpectralTwoPointUnitary}. The fact that the present model is scaled to the vicinity of the edge only makes its appearance in the
scaling variable, $s=\pi \omega /\Delta$, which now makes reference to the near edge
level spacing, $\Delta=\rho^{-1}=\pi e^{-S_0}\zeta^{-1/2}$. 

Before concluding this section, let us comment on a point that will become relevant in connection with a minimal string theory interpretation in the next section. We derived the representation Eq.~\eqref{eq.SuperKontsevich} by fluctuation expansion of the action around $\zeta=0$, right at the edge where the mean field symmetry breaking comes to an end by definition. On the other hand, we know that for any $\zeta>0$ mean field causal symmetry breaking  is equivalent to the statement of a finite density of states on average. We can invoke this mechanism by a `second' stationary phase approximation, this time applied to the fluctuation action Eq.~\eqref{eq.SuperKontsevich}, i.e. by asking for the least action fluctuation configurations, $\bar a$, for given $\zeta>0$. A straightforward variation leads to 
\begin{align}
    \label{eq.KontsevichSaddle}
    \bar a= \pm i\sqrt{\zeta},
\end{align}
and the differentiation in sources at the causal of these yields the mean field spectral density in agreement with Eq.~\eqref{eq.SpectralDensityEdge}. Comparing with the exact result, Eq.~\eqref{eq.AiryDos}, we interpret the mean field result as the approximation of the Airy functions in the semiclassical limit $e^{S_0}\gg 1$.

\subsubsection{Disks, annuli and instantons}
Having described the physics of spectral branes in terms of a double-scaled matrix
model, we would now like to make the connection to the continuum approach of minimal
string theory, which serves as the bulk description in this setting. We start by
discussing a single determinant operator insertion. As described in
\cite{Seiberg:2003nm}, in the bulk the density of states is related to a disk
amplitude with boundary condition given by a single energy variable $\zeta$, that is
a worldsheet with a single FZZT brane boundary condition $\zeta$, referred to in this
context as the boundary cosmological constant. The connection between determinant operator insertion and spectral density is made by a reduced version of the flavor matrix  
integral \eqref{eq.SuperKontsevich} 
\begin{align}
\label{eq.AirhBrane}
\Psi(\zeta):=\left\langle {\rm det}(\zeta-H)\right\rangle = \int da\, e^{-e^{S_0}\left(\frac{a^3}{3} + \zeta a \right)} =   e^{-S_0/3} 2\pi i{\rm Ai}\left( -e^{2S_0/3}\zeta \right)
\end{align}
where $a, \zeta$ are single scalar variables (the former representing a one-dimensional flavor matrix) and in the last equality we noted that for an appropriate choice of integration contour the integral  takes the form of a well-known integral representation of the Airy
function (see Appendix~\ref{app.Airy} for details.). We interpret this representation as the effective
theory of an FZZT brane upon integrating out the contribution of the ZZ branes,
in terms of a single  degree of freedom $a$.

At the end of the previous section we have seen that 
in the semiclassical evaluation of this miniature flavor theory we encounter two saddles,  Eq.~\eqref{eq.KontsevichSaddle}, a toy model version of the standard saddle and the Altshuler-Andreev saddle. Summation over these saddles and the fluctuations around them produced a large $\zeta$ semiclassical approximation of the spectral determinant, and the full $a$-integral an restoration of the exact result Eq.~\eqref{eq.AirhBrane}. Translating to the bulk, the double saddle point structure means that the  brane exhibits two branches, where the
 second appears on the brane as an instanton. In the literature\cite{Maldacena:2004sn}, this saddle point structure has been interpreted as a semiclassical approximation to the  exact quantum target space of the brane, with $\zeta$ playing the role of an effective semiclassical target space coordinate.

So far we only considered a single determinant, which is not rich enough
to exhibit Goldstone modes associated to symmetry breaking.  In order to get from the toy model to the full
problem, we need to consider more than one FZZT brane. From the perspective of this paper, the most natural realization would be one corresponding to the ratio of spectral determinants, with an effective supersymmetric flavor matrix representation.  However, absent a matching bulk picture\footnote{This would require a more
explicit microscopic understanding of the ghost branes of \cite{Okuda:2006fb} in general, and specifically in terms of the continuum worldsheet approach of minimal string theory. In view of how much simpler and transparent  the non-perturbative analysis
becomes compared to the replica approach, this might be a rewarding investment.} we engage the replica formalism whose starting point are $R$-fold replicated determinants $\left\langle
\det(\zeta_1-H)^R \det(\zeta_2-H)^R \right\rangle$. 
The matrix theory implementation of these products in terms of $R$-fold replicated
Grassmann integrals is straightforward. The construction of
section~\ref{sec:RMTSigmaModel} modified for $2R$-Grassmann integration variables and
zero commuting ones leads to a flavor matrix model \eqref{eq.MinimalFlavorAction}
where $a$ is now a $2R\times 2R$ matrix. In the double scaled limit, this leads to
\eqref{eq.SuperKontsevich} from where the non-perturbative structure of spectral
correlations is obtained via a replica version of the AA saddle point detailed in
Appendix ~\ref{sec:supersymmetry_vs_replicas}. 

An alternative approach, more closely geared towards a bulk interpretation, starts from the representation of the double scaled expectation value, \cite{Morozov:1994hh,Maldacena:2004sn}
\be\label{eq.MultiDeterminantBulk}
\left\langle \prod_{i=1}^{R}\det (\zeta_i-H)  \right\rangle  = \frac{\Delta(d)}{\Delta(\zeta)}  \prod_{i=1}^{R} \Psi(\zeta_i)\,,
\ee
where the `wavefunction' $\Psi(\zeta)$ is defined in Eq.
\eqref{eq.AirhBrane} above. Here $\Delta(\zeta) = \prod_{i<j} (\zeta_i - \zeta_j)$ is the
Vandermonde determinant, and similarly for $\Delta(d)$, where $d=e^{-S_0}(\partial_1,\dots,\partial_R)$ contains derivatives
 acting  $i^{\rm th}$ coordinate $\zeta_i$ of the wavefunction $\Psi(\zeta_i)$. In continuum language, the
individual contributions in \eqref{eq.MultiDeterminantBulk} are associated to
particular worldsheet contributions. In the semiclassical limit $e^{S_0}\gg 1$, the product $\prod_{i=1}^{R} \Psi(\zeta_i)$ becomes a
product of WKB wavefunctions, each factor comprising a disk diagram as well as an annulus
with both boundaries on the same brane
\cite{Maldacena:2004sn},  and the ratio $\frac{\Delta(d)}{\Delta(\zeta)}$ comes from
exponentiated annuli with boundary conditions on different branes. Let us see explicitly how this works for the case of two determinants, that is two exponentiated loop operators \eqref{eq.LoopOperator},
\be
\left\langle \Psi(\zeta_1) \Psi(\zeta_2)  \right\rangle=\left\langle e^{{\cal W} (\zeta_1)} e^{{\cal W}(\zeta_2)}\right\rangle\,.
\ee
A single insertion of the loop operator is the matrix-model representation of a worldsheet with a single boundary, the so-called disk amplitude, Disk$(\zeta)=\langle {\cal W}(\zeta) \rangle$. A two-loop correlator gives the annulus diagram ann$(\zeta_1,\zeta_2) = \langle {\cal W}(\zeta_1) {\cal W}(\zeta_2) \rangle$. Then to leading semiclassical order we can write, \cite{Saad:2019lba,Blommaert:2019wfy},
\be\label{eq.diskAnnulus}
\left\langle \Psi(\zeta_1)  \Psi(\zeta_2) \right\rangle \simeq e^{  {\rm Disk}(\zeta_1) +  {\rm Disk}(\zeta_2) + {\rm ann}(\zeta_1,\zeta_2) + \frac{1}{2}\left( {\rm ann}(\zeta_1,\zeta_1) + {\rm ann}(\zeta_2,\zeta_2) \right)  }
\ee
We note that the individual factors contributing to this answer are precisely the ingredients we introduced above, namely disk diagrams with a boundary on either of the spectral branes, Disk$(\zeta_{1,2})$ annulus diagrams with both boundaries on the same brane, ann$(\zeta_{1,2},\zeta_{1,2})$, and finally an annulus diagram with one boundary on each spectral brane, ann$(\zeta_1,\zeta_2)$.

In the next section, we will analyse the connected annular contribution ann$(\zeta_1,\zeta_2)$ and extract the singular (in
energy) terms coming from the EFT predictions in Section \ref{sec.EFTofChaos} above.

\subsection{Universal RMT contributions in 2D gravity}
\label{sec.ContinuumGravityFormulation}

In this section, we show how universal
contributions of the EFT of chaos in Section \ref{sec.TopologicaExpansionEFT} are recovered from
the bulk perspective. To this end we evaluate the worldsheet diagrams contributing to
\eqref{eq.MultiDeterminantBulk} in the replica limit $R\to 0$ and match them to individual contributions
in Section \ref{sec.EFTofChaos}. Carrying out the disk and annulus expansion, \eqref{eq.diskAnnulus} for the replicated determinants, one finds
\bea
W(\zeta_1,\zeta_2)&=&\partial^2_{\zeta_1,\zeta_2}\lim_{R\rightarrow 0} \frac{\left(\Psi(\zeta_1)^R -1\right) \left(\Psi(\zeta_2)^R -1\right)}{R^2}\nonumber\\
& =& \partial^2_{\zeta_1,\zeta_2} {\rm ann}(\zeta_1,\zeta_2)
\eea
i.e. the only surviving connected contribution to the correlator of two resolvents, \eqref{eq.TraceResolvent}, is indeed the annulus ann$(\zeta_1,\zeta_2)$. This result is obvious once we recall that one can find the two-resolvent expectation value simply by computing the expectation value of two loop operators, $\langle {\cal W}(\zeta_1) {\cal W}(\zeta_2) \rangle$, which is nothing but our earlier definition of ann$(\zeta_1,\zeta_2)$ (see Figure \ref{fig.MMSTannulus}). However it is reassuring to see that it also follows from the full machinery developed above. In particular, it would be impossible to develop our causal symmetry breaking analysis without full recourse to the determinants \eqref{eq.MultiDeterminantBulk}. We have thus established that we can extract the leading RMT singularity $\sim s^{-2}$ in bulk language from an annulus diagram with two FZZT boundaries, $Z(\zeta_1,\zeta_2)$,
the so-called FZZT annulus partition function in minimal string theory.   This amounts to computing the contribution of the Goldstone modes to the leading singularity around the standard saddle in the bulk, i.e. the exact bulk equivalent of the EFT diagram \eqref{eq.GoldstoneCylinder}.

We thus want to compute the bulk annulus diagram, which in the worldsheet setting  amounts to 
adding  contributions from worldsheet ghosts $Z_{\rm ghost}$, the 2d Liouville
partition function $Z_{\rm Liouville}$ as well as the $(2,p)$ minimal model CFT,
$Z_{\rm matter}$, leading to the final result,
\begin{align}\label{eq.FZZTAnnulusAmplitude}
Z_{(2,p)}(\zeta_1,\zeta_2) = \int_0^\infty dP \frac{\cos(4\pi s_1 P)  \cos(4\pi s_2 P)}{2\pi P \sinh 2\pi \frac{P}{b}  \cosh 2\pi \frac{P}{b}}\,,
\end{align}
which is the $(2,p)$ version of what we called ann$(\zeta_1,\zeta_2)$ above, and where
\begin{align}
\label{eq.MinimalStringDefinitions}
\zeta_{1,2} =\sqrt{2\pi} \kappa \cosh 2 \pi bs_{1,2}\,,\qquad b = \sqrt{\frac{2}{p}}\,,\qquad \kappa = \sqrt{\frac{\mu}{\sin^2 \pi b^2}}\,.
\end{align}
The constant $\mu$ is the bulk cosmological constant of 2D gravity and we have chosen the normalization of the energy variables $\zeta_{1,2}$ so as to reproduce the leading density of states \eqref{eq.SpectralDensityEdge}. A detailed exposition of the relevant minimal-string calculations can be consulted in \cite{Fateev:2000ik,Martinec:2003ka,Mertens:2020hbs}.
\subsubsection{The ramp}
\begin{figure}[t]
\begin{center}
\includegraphics[width=.65\columnwidth]{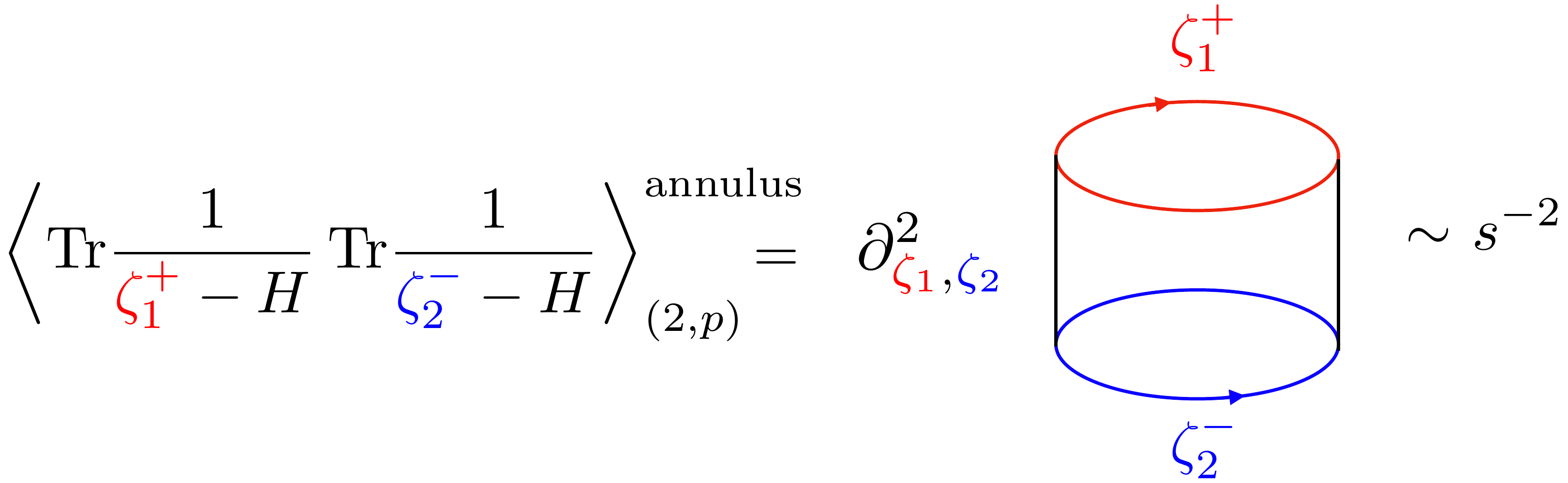}\\
\caption{The annulus contribution to the spectral two-point function. Its universal part gives the correct EFT singular contribution of $-1/2s^2$. The diagram is identical to \eqref{eq.GoldstoneCylinder}, although here it represents a minimal-string worldsheet interpreted as a 2D spacetime with two branes, also known as the Laplace transform of a Euclidean wormhole. The 2D gravity amplitude gives the same leading singularity including the numerical coefficient as \eqref{eq.GoldstoneCylinder}, coming from the Goldstone sector of causal symmetry. \label{fig.MMSTannulus}}
\end{center}
\end{figure}
 The first non-trivial contribution to the spectral correlation function is obtained by differentiating the FZZT annulus partition function, \eqref{eq.FZZTAnnulusAmplitude}, twice with respect to the energy arguments
 \be
W^{\rm annulus}_{(2,p)} (\zeta_1,\zeta_2) =  \partial^2_{\zeta_1,\zeta_2}Z_{(2,p)}(\zeta_1,\zeta_2)\,,
 \ee
 which as we explained above is just the two-resolvent correlation function, now written in 2D gravity language (cf. Fig.~\ref{fig.MMSTannulus}).  The differentiation with respect to energy arguments proceeds implicitly, using \eqref{eq.MinimalStringDefinitions} and the resulting integral can be carried out explicitly, with the answer, \cite{Martinec:2003ka,Seiberg:2003nm,Mertens:2020hbs},
 \bea
W^{\rm annulus}_{(2,p)} (\zeta_1,\zeta_2) = \frac{{\rm sech}(b\pi s_1) {\rm sech}(b \pi s_2)}{32 \pi \kappa^2 \left( \cosh(\pi b s_1) + \cosh(\pi b s_2)  \right)^2}\,.
\eea
 This finally gives the simple expression,

\bea\label{eq.MMSTLeadingSingularDiagram}
W^{\rm annulus}_{(2,p)} (\zeta_1,\zeta_2) = \frac{1}{4}\frac{1}{\sqrt{-\zeta_1 + \kappa} \sqrt{-\zeta_2 + \kappa}}  \frac{1}{(\sqrt{-\zeta_1 + \kappa}+\sqrt{-\zeta_2 + \kappa})^2}\,.
\eea
The spectral correlation function is then obtained via
\bea
R_2(\zeta_1 - \zeta_2) &=& \frac{\Delta^2}{2\pi^2}{\rm Re} \,W(\zeta_1^+,\zeta_2^-)_{\rm annulus} + {\rm reg.} \nonumber\\
&=& - \frac{\Delta^2}{2 \pi^2 (\zeta_1-\zeta_2)^2}=- \frac{1}{2s^2}\,.
\eea
 The terms we left out are regular in the limit $\zeta_1 \rightarrow \zeta_2$, as
 indicated in the first line. Taking the real part is equivalent to fixing a prescription to analytically continue the function $W(\zeta_1,\zeta_2)_{\rm annulus}$, which leads to the singular behavior we indicated. This is the bulk counterpart of the cylinder
 contribution in the topological expansion of the EFT of chaos given in Eq.
 \eqref{eq.GoldstoneCylinder} and it is reassuring that the singular part of the
 diagram indeed coincides with that prediction. In this way it becomes  clear
 how to map the advanced (red) and retarded (blue) boundaries in that representation to the properties of
 the FZZT boundary state given by the boundary cosmological constants
 $\zeta_{1,2}^\pm$ (see Figure \ref{fig.MMSTannulus}). 

While higher-genus contributions, such as
 \eqref{eq.TwoCrossesOrientable} in symmetry classes different from the unitary one might be possible to compute in principle (but difficult in
 practice),  the bulk interpretation of the Althshuler-Andreev saddle remains elusive, although it seems promising to attempt to construct these contributions from applying the  Riemann-Siegel lookalike construction to contributions like \eqref{eq.MMSTLeadingSingularDiagram}. This would amount to giving a bulk semiclassical description of the leading singularity around the Altshuler-Andreev saddle, which we discussed from the EFT perspective in \eqref{eq.TheFullStructure}. The same is true for the physics of spectral correlations in the far infrared, where the level rigidity symptomatic for systems with hard quantum chaos reflects in a restoration of causal symmetry and integration over the full flavor matrix manifold. Since the non-perturbative structure of the EFT is most accessible in the supersymmetric formulation, it would be desirable to study the extension of the string worldsheet formalism to one that is dual to the supersymmetric (i.e. graded) flavor framework.

\subsection{A three-dimensional example: $\mathbb{T}^2\times I$ wormholes}
In order to further illustrate the universality of the EFT, let us briefly highlight how the leading $1/s^2$ singularity appears in three dimensional gravity. The relevant computations are described in references by Cotler and Jensen, \cite{Cotler:2020ugk,cotler2020ads3}, who give a path-integral quantization of three dimensional gravity on spacetimes with the topology of a torus times an interval, $\mathbb{T}^2\times I$. This is the 3D analogue of the annular geometry in the minimal string theory above and we therefore expect to be able to extract the information of the EFT diagram \eqref{eq.GoldstoneCylinder} from this. We start with the expression of the two-boundary partition function, \cite{Cotler:2020ugk},
\be
Z_{\mathbb{T}^2\times I}(\tau_1,\tau_2) = \frac{1}{2\pi^2} Z_0(\tau_1)Z_0(\tau_2) \sum_{\gamma \in {\rm PSL}(2;\mathbb{Z})} \frac{\Im (\tau_1)\Im (\tau_2)}{|\tau_1 + \gamma \tau_2|^2}
\ee
where $\tau_1$ and $\tau_2$ are modular parameters of the two tori, $\gamma$ is a M\"obius transformation on $\tau_2$ and $Z_0(\tau) = \frac{1}{\sqrt{\Im \tau}|\eta(\tau)|^2}$ in terms of the Dedekind eta function. One can show that the leading low-temperature behavior at fixed spins $s_{1,2}$ on each of the boundaries gives,  \cite{Cotler:2020ugk}
\be
Z_{\mathbb{T}^2\times I}(\beta_1,\beta_2) \sim \frac{1}{2\pi}\frac{\sqrt{\beta_1 \beta_2}}{\beta_1 + \beta_2} e^{-E_1 \beta_1 - E_2 \beta_2} + \cdots
\ee
This expression can be analytically continued to real time by setting $\beta_{1,2}\rightarrow \beta \pm i T$, which was used in \cite{Cotler:2020ugk} to obtain the linear ramp behavior. Since we are interested in the partition function at fixed energies $E_{1,2}$ instead, we first Laplace transform this expression with respect to the two temperature parameters, resulting in
\be
Z_{\mathbb{T}^2\times I}(E_1,E_2) =  \frac{1}{2\pi}\int_0^\infty  \frac{\sqrt{\beta_1 \beta_2}}{\beta_1 + \beta_2} e^{-E_1 \beta_1 - E_2 \beta_2} = \frac{1}{4 \sqrt{E_1}\sqrt{E_2} (\sqrt{E_1} + \sqrt{E_2})^2}
\ee
Note the structural similarity with the expression \eqref{eq.MMSTLeadingSingularDiagram} above. This allows us to immediately deduce the contribution of this diagram to the spectral two point function
\bea
R_2(\omega) &=& \frac{\Delta^2}{2\pi^2}{\rm Re} \,Z_{\mathbb{T}^2\times I}(E_1^+,E_2^-)+ {\rm reg.} \nonumber\\
&=&- \frac{\Delta^2}{2\pi^2 (E_1-E_2)^2}= -\frac{1}{2s^2}
\eea
which  gives the universal ramp contribution to the spectral form factor predicted by the Goldstone diagram \eqref{eq.GoldstoneCylinder}. Again, the question of how to extend the bulk result to the non-perturbative level (i.e. ${\cal O}(e^{-L})$) remains open.

\section{Discussion}\label{sec.Discussion}
The main theme in this paper has been the universal nature of chaotic spectral
correlations governed by a symmetry-based effective field theory. One of the
attractive features of this approach is that it applies both to ensembles, such as
random matrix theory with an arbitrary invariant potential, or disorder-type models,
such as the SYK model, but crucially also to individual theories without  reference
to ensembles or disorder averages. In the latter case the content of the effective
field theory is to be understood as applying to the individual quantum system by
providing an envelope function of the typical behavior of connected correlation
functions or similar observables, where the true correlation function fluctuates
potentially quite wildly around the envelope. The exact envelope behavior can be
recovered by some `mild' averaging, say a running time average of a real-time
correlation function or an average over judiciously chosen energy intervals for a
momentum-space correlation function. At the same time we have shown that specific
gravity configurations (e.g. two-sided Euclidean wormholes) allow us to compute the
universal enveloping function from the gravity perspective. One may thus entertain
the hope that similar bulk solutions can likewise be associated to the computation of
connected spectral correlations in higher-dimensional AdS/CFT pairs, such as the ABJM
theory \cite{Aharony:2008ug} or ${\cal N}=4$ SYM theory \cite{Maldacena:1997re}
without having to construct ensembles of boundary theories (other than perhaps via
some 'mild' averaging over a small set of parameters, e.g. moduli or coupling constants). In this way the
EFT perspective gives a quite general justification to consider bulk wormhole-type
contributions even for individual boundary theories, as would be desirable from the
perspective of the recent bulk computations of the unitary Page curve from `replica
wormholes' \cite{Penington:2019kki,Almheiri:2019qdq,Marolf:2020xie,Almheiri:2020cfm}
(see also  \cite{Liu:2020rrn} for a review emphasizing the many-body aspects as
relevant to this work).

The idea that wormhole-type correlations are to be associated with the universal behavior of quantum chaotic systems has been investigated from the perspective of the eigenstate thermalization hypothesis  in \cite{Pollack:2020gfa} and from the point of view of a conjecture of the statistics of the operator-product coefficients in \cite{Belin:2020hea}. It would be interesting to investigate further the relation between these approaches and ours. For this it may be useful to note that FZZT boundary conditions of the type discussed in Section \ref{sec.CausalSymmetryMMST} also play a role in establishing eigenstate thermalization in simple holographic models \cite{Nayak:2019evx}. While it is typically hard to establish that a given theory or class of theories satisfies the eigenstate thermalization (the recent results of \cite{Sonner:2017hxc,Nayak:2019khe,Nayak:2019evx}, as well as \cite{Saad:2019pqd} from the gravity perspective notwithstanding), we have established here results of a morally very similar nature, which are however of a wide -- in fact in a technical sense universal -- applicability.

In cases where the effective field theory came from an underlying individual quantum system there is a sense in which correlation functions of the `multi-boundary' type should refactorize into individual `single-boundary' contributions, and it would be interesting to investigate how this can be incorporated into the EFT of chaotic quantum systems. Progress in this direction in JT gravity as well as minimal string theory, albeit without recourse to causal symmetry, was presented recently in \cite{Blommaert:2019wfy,Blommaert:2020seb}.

It would be an important goal to examine higher-dimensional examples, and in particular to compare with recent attempts to associate ensemble-type interpretations to the bulk path integral. Our symmetry-based approach gives a clear picture in what sense it is appropriate to interpret fixed individual quantum systems with ensembles. Using the classification result of Refs.\cite{Altland1997,Heinzner2005} together with the EFT Lagrangian \eqref{eq.EFTactionInTermsofvF} adapted to the correct symmetry class gives a direct algorithm to identify and write down the appropriate ensemble for any boundary theory of interest. Since the higher-dimensional bulk story in full generality seems to require a third quantized "Universe Field Theory", (see discussions in \cite{Giddings:1988wv,Marolf:2020xie,Anous:2020lka,Chen:2020tes}), a more modest intermediate goal might be to calculate the coefficients of universal singular diagrams predicted by our EFT from individual bulk manifolds, as we have done here for two and three-dimensional gravity.

\subsection*{Acknowledgements}
We would like to thank Alexandre Belin, Jan de Boer, Jie Gu, Thomas Guhr, Daniel Jafferis, Marcos Mari\~no, Pranjal Nayak, Francesco Riva, Steven Shenker, Douglas Stanford, Manuel Vielma for enlightening discussions regarding the contents of this paper, and especially Pranjal Nayak and Manuel Vielma for ongoing collaborations on related matters. This work has been supported in part by the Fonds National Suisse de la Recherche Scientifique (Schweizerischer Nationalfonds zur F\"orderung der wissenschaftlichen Forschung) through Project Grants 200020\_ 182513, the NCCR 51NF40-141869 The Mathematics of Physics (SwissMAP), and by the DFG Collaborative Research Center (CRC) 183 Project No. 277101999 - project A03.

\appendix

\section{Expansion in Goldstone modes vs. topological recursion} 
\label{ssub:expansion_in_goldstone_modes_vs_topological_recursion}
\begin{figure}[h]
\begin{center}
\includegraphics[width=.6\columnwidth]{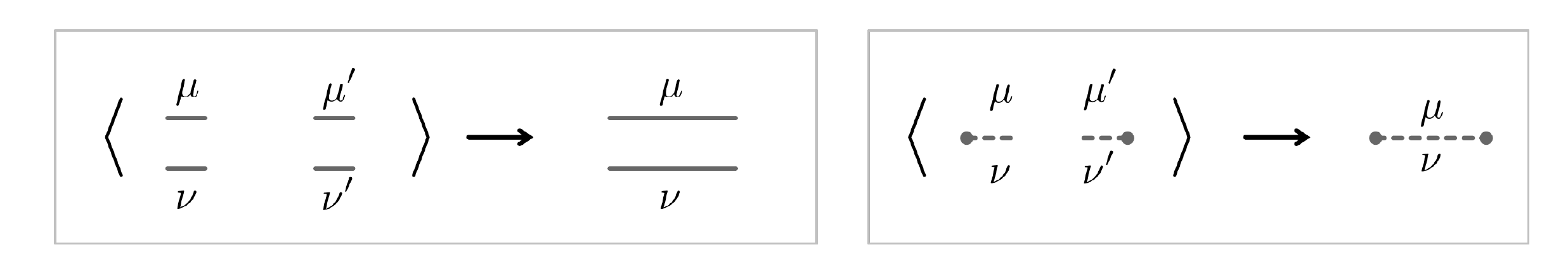}\\
\end{center}
\caption{Alternative ways of representing building blocks of matrix theories. Left: double line notation customary in matrix field theory. Right: `impurity diagram representation' customary in the physics of random systems. }
 \label{fig:DoubleVsDashedLine}
\end{figure}

As mentioned in the main text, the sigma model approach is applicable to a wide class of chaotic Hamiltonians, including to matrix Hamiltonians $H=\{H_{\mu\nu}\}$ drawn from distributions $P(H)dH$. Referring to section~\ref{sec:RMTSigmaModel} for a technically detailed discussion, we here stay on a qualitative level and compare the EFT approach to the topological (genus) expansions central to traditional matrix field theory. It is instructive to formulate this comparison in a diagrammatic language, where we will us two slightly different languages for the representation of the theory's microscopic building blocks depicted in Fig.~\ref{fig:DoubleVsDashedLine}. On the left, we have the double line notation standard in matrix field theory, where individual matrix elements $H_{\mu\nu}$ are represented by parallel stubs, and their Wick contraction is indicated by a double line. The diagrammatic codes standard in the theory of disordered systems --- which are equivalent but arguably are more efficient in conveying the information relevant to us --- the same contraction is indicated by a single dashed line, as indicated on the right.

\FloatBarrier

\begin{figure}[h]
\begin{center}
\includegraphics[width=.6\columnwidth]{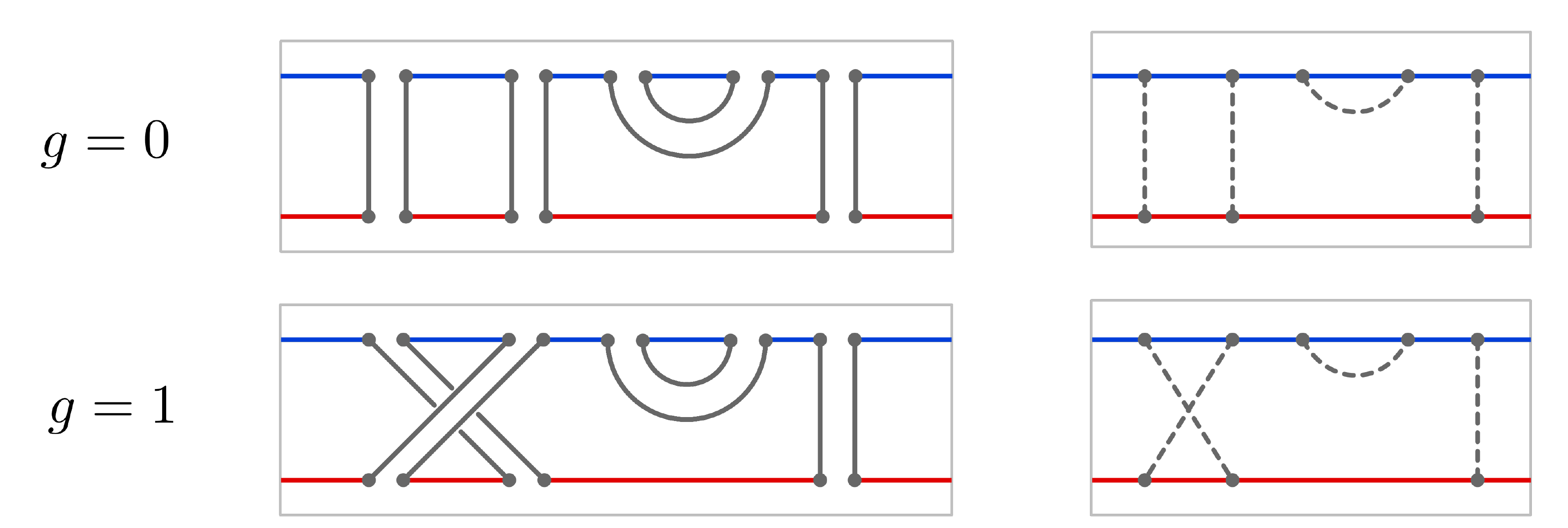}\\
\end{center}
\caption{Alternative ways of representing building blocks of matrix theories. Left: double line notation customary in matrix field theory. Right: `impurity diagram representation' customary in the physics of random systems. }
 \label{fig:PlanarVsNonPlanar}
\end{figure}

Presently, these building blocks appear in the context of the advanced and retarded resolvents $G(z^+)=(z^+-H)^{-1}$ and $G(z^{\prime -})=(z^{\prime -}-H)^{-1}$, which we may consider formally expanded in $H$. Upon averaging, this then leads to structures as indicated in Fig.~\ref{fig:PlanarVsNonPlanar} in double line (left) or impurity (right) syntax. Each diagram contributes to the perturbative expansion with a factor $\sim L^{-g}$, where in matrix theory parlance the power $g$ measures its degree of non-planarity, or genus  (cf. the upper and lower left panel), and in that of condensed matter theory  the number of crossing impurity lines (upper and lower right panel).

\FloatBarrier

\begin{figure}[t!]
\begin{center}
\includegraphics[width=.9\textwidth]{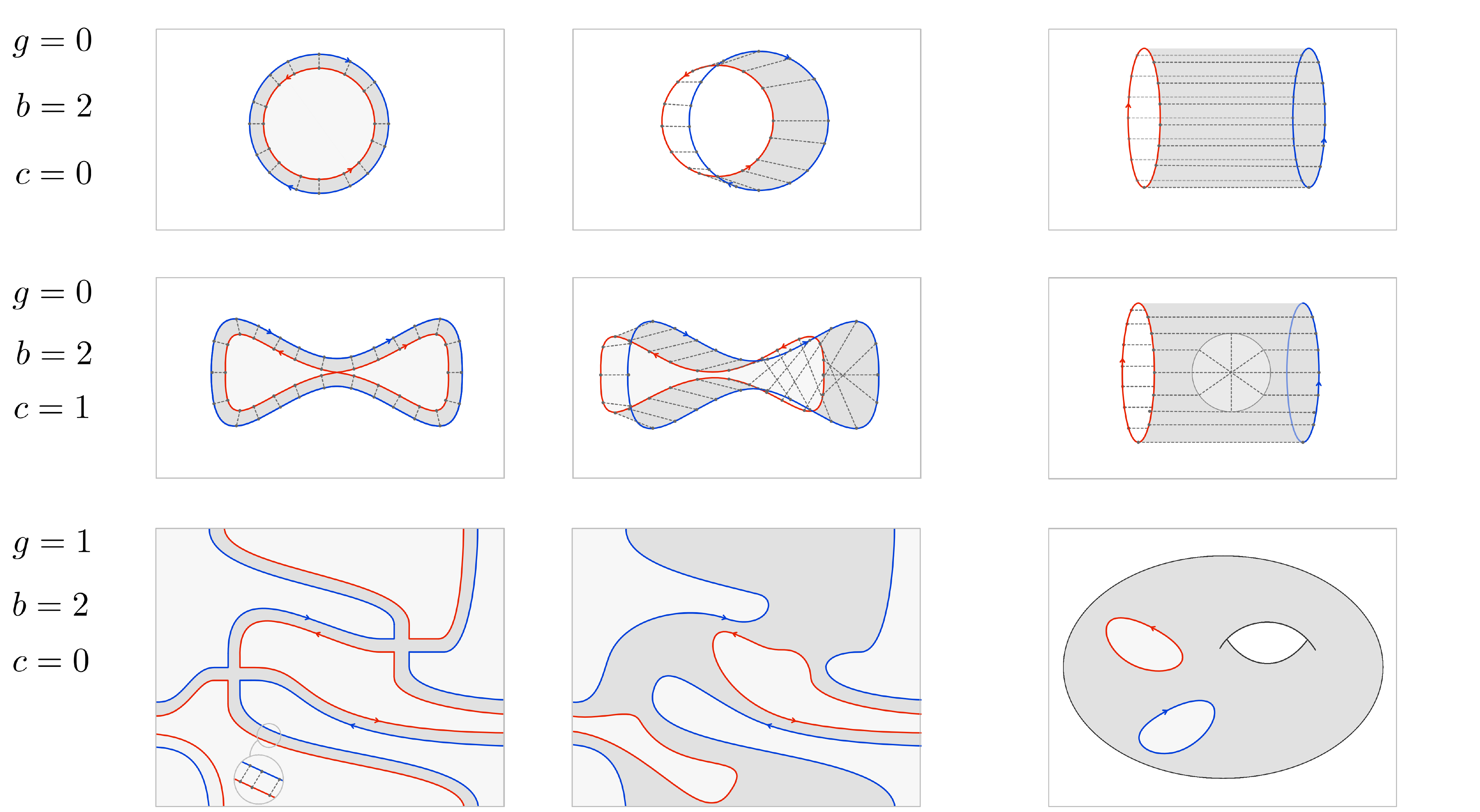}\\
\end{center}
\caption{Impurity diagram representation of the perturbative surface to bulk correspondence. Each line is labelled by the genus $g$, the number of boundaries $b$ and the number of crosscaps $c$. Discussion, see text.}
 \label{fig:ImpurityStrings}
\end{figure}

For certain observables, topological recursion formulae establish
quantitative connections between the total contribution of diagrams of genus $g+1$ to
that of one degree lower, $g$. This machinery has been engaged in
Ref.~\cite{Saad:2019lba} to establish a boundary--bulk correspondence by identifying
equivalent hierarchies in the topological expansion of  JT gravity purportedly
equivalent to a matrix integral at the boundary. The perturbative approach to chaos or disorder in condensed matter physics is somewhat different: Its starting point is the observation that a tiny (exponential in $L$) subset of all diagrams contributing to a specific genus class is capable of producing a singularity in $\omega^{-g}$, where $\omega=|z-z'|$. We emphasize that the degree of the most violent singularity is precisely matched by the genus, so that the effective perturbative parameter of a series retaining only the maximally singular diagrams reads $1/(N\omega)^{g+l}\sim 1/s^{g+l}$, where $l$ is fixed and depends on the specific definition of the observables. (For example, $l=2$ in the case of the spectral two point correlation function.) This is the expansion parameter featuring in the expansion of the EFT, indicating that the singular diagrams afford an interpretation in terms of Goldstone mode propagators and vertices of the field theory. To see this in more explicit ways, consider the three examples of singular diagrams contributing to the spectral correlation function (the closed fermion loops representing the trace over resolvents) shown in the left column of Fig.~\ref{fig:ImpurityStrings} 

\FloatBarrier

The distinguishing feature of these diagrams is that they (a) define a subset of much
higher order (lower entropy) compared to the full contents of a given genus class,
(b) can be computed in closed form for matrix models or other `microscopically'
defined models of sufficient simplicity, and universally yield equal
contributions to observables. For example, the figure $\infty$ diagram
featuring in the second row appears in random matrix theory, the periodic orbit
approach to quantum billiards (the Sieber-Richter contribution~\cite{Sieber_2001}),
and the theory of disordered metals (weak localization correction, see Ref.~\cite{Larkin1980} for a historic reference) as a contribution
of identical topology but differently defined microscopic details. The actual computation of the diagrams from first principles can be tricky, equivalent in complexity to the construction of the EFT from a microscopically defined model (we will sidestep this complication for the time being.) However, in the deep infrared, they all contribute as $\sim s^{-(l+g)}$ with universal coefficients, and this 
topological equivalence of the most singular  contributions
across different theories is the key to understanding random matrix universality from
semiclassical principles~\cite{Muller2005}.

The EFT approach automatically filters out the singular diagrams and represents their contribution in terms of Goldstone mode fluctuations as discussed above. However, it is instructive to stay for a moment on the microscopically resolved level of a matrix theory and interpret the correspondence indicated on topological grounds in Eqs.~\eqref{eq.GoldstoneCylinder} to~\eqref{eq.TwoCrossesOrientable} in the microscopically resolved language. To this end, consider the three
representatives of singular `impurity diagrams' shown in the first column of the
bottom part of the figure. Here, the first (`wagon wheel') has $g=0$ and accordingly is singular in $s^{-2}$,
translating to the linear in $\tau$ ramp upon Fourier transformation. (For later
reference we note that the wheel comes in two incarnations, one with arrows in
opposite orientation, and another, not shown and existing only in systems with time
reversal invariance, with arrows in identical orientation.) Now pull the two rings
forming the wheel apart to generate a cylinder covered with a pattern of parallel
scattering lines. The latter defines a bulk surface describing the wheel shaped ribbon in different representation. 

For a less trivial example, consider the $g=1$ diagram which contains  stretches of mutually parallel and
anti-parallel propagator lines  and hence exists only in systems with time reversal symmetry.
Its higher loop order reflects in a stronger singularity, $\sim s^{-3}$, yielding a
sub-leading correction $\sim \tau^2$ to the form factor of systems in the
corresponding symmetry classes. This time, the rearrangement into two non-crossing
propagator loops requires a section of maximally crossed statistical lines, which is
information equivalent to the presence of a cross-cap in surface representation.
Finally, the third diagram is one of two representatives with $s^{-4}$ singularity,
mutually canceling each other in their contribution to the spectral form factor but
not to other observables. While a consistently opposite arrow orientation indicates
that time reversal symmetry is not essential, a representation avoiding the crossing
of propagator lines requires an underlying torus topology (indicated by identified
boundaries in the figure.) Again we may pull the propagator lines apart to end up
with two loops connected by a surface that contains a handle and is covered with a
fabric of non-intersecting scattering lines.

The high level of universality manifest in the topology of the diagrammatic expansion suggests the presence of a simple underlying principle. The latter becomes obvious once we return to the level of the EFT and abandon the perturbative approach in favor of a non-perturbative one.

\section{Supersymmetry vs replicas} 
\label{sec:supersymmetry_vs_replicas}

Supersymmetry is the method of choice when it comes to the investigation of non-perturbative structures in quantum chaos. However, occasionally, supersymmetry is not an option and we must resort to the alternative formalism of replicas instead. Presently this happens in connection to our discussion of minimal string theory, where formulations dual to a matrix theory containing inverse determinants (`ghosts') are not available. We therefore must do with a fermionic, non-graded version of the theory.

While in the early days of the field replicas have been dismissed as ill suited to the extraction of any non-perturbative information~\cite{Verbaarschot_1985}, it later became clear that this verdict had been overly harsh. Beginning with Ref.~\cite{Kamenev1999}, various non-perturbative effects in spectra and wave function localization have been successfully described by replica methods. To get the idea and touch base with the supersymmetry formalism, consider the product $\mathcal{Z}\equiv \prod_{i=1}^R \det(z_i-H)$ of $R$ determinants. Employing the first of $R$ energy arguments as a source, we obtain
\begin{align}
    \partial_{z_1}\mathcal{Z}\big|_{z_i=z}=\partial_{z_1}|_{z_1=z}\det(z_1-H) \det(z-H)^{R-1}\stackrel{R\to 0}\longrightarrow \frac{\partial_{z}\det(z-H)}{\det(z-H)}.
\end{align}
This expression teaches us that in the replica limit $R\to 0$ the fermionic partition
sum assumes a form identical to that of the supersymmetric one
Eq.~\eqref{eq.Z2Intro}, where the $R-1\to -1$ fermionic spectator determinants turn
into a `missing fermionic determinant' assuming a role of an bosonic one. Of course,
all this presumes that the replica limit is well defined. Pushing on and
subjecting the determinants to the field theory machinery we obtain an $R$-flavor
theory, or $2R$ flavors if retarded and advanced resolvents are correlated with each
other. At the saddle point level, we accordingly encounter $2^{2R}$ different choices
of saddle point solutions. However, in view of the above fermion-boson analogy, one
expects that only one of these assumes the role of the AA saddle. This is the one,
where in each of the $2(R-1)$ spectators we pick the `causal saddle', and in the
sourced replica channel no. 1 invert the sign. Formally, one indeed finds that in the
replica limit, the fluctuation contributions around all other saddles vanish. Adding
the surviving replica AA saddle to the standard one, the correlation
function~\eqref{eq.SpectralTwoPointUnitary} is recovered~\cite{Kamenev1999}. The recovery of the full result by saddle point analysis is again owed to the semiclassical exactness of the integrand. However, even for symmetry classes were this not fulfilling this criterion, an asymptotic expansion around the two saddle points yields excellent descriptions of spectral correlations to any desired precision.

\section{Airy function}\label{app.Airy}
\begin{figure}[t]
\begin{center}
\includegraphics[width=.5\columnwidth]{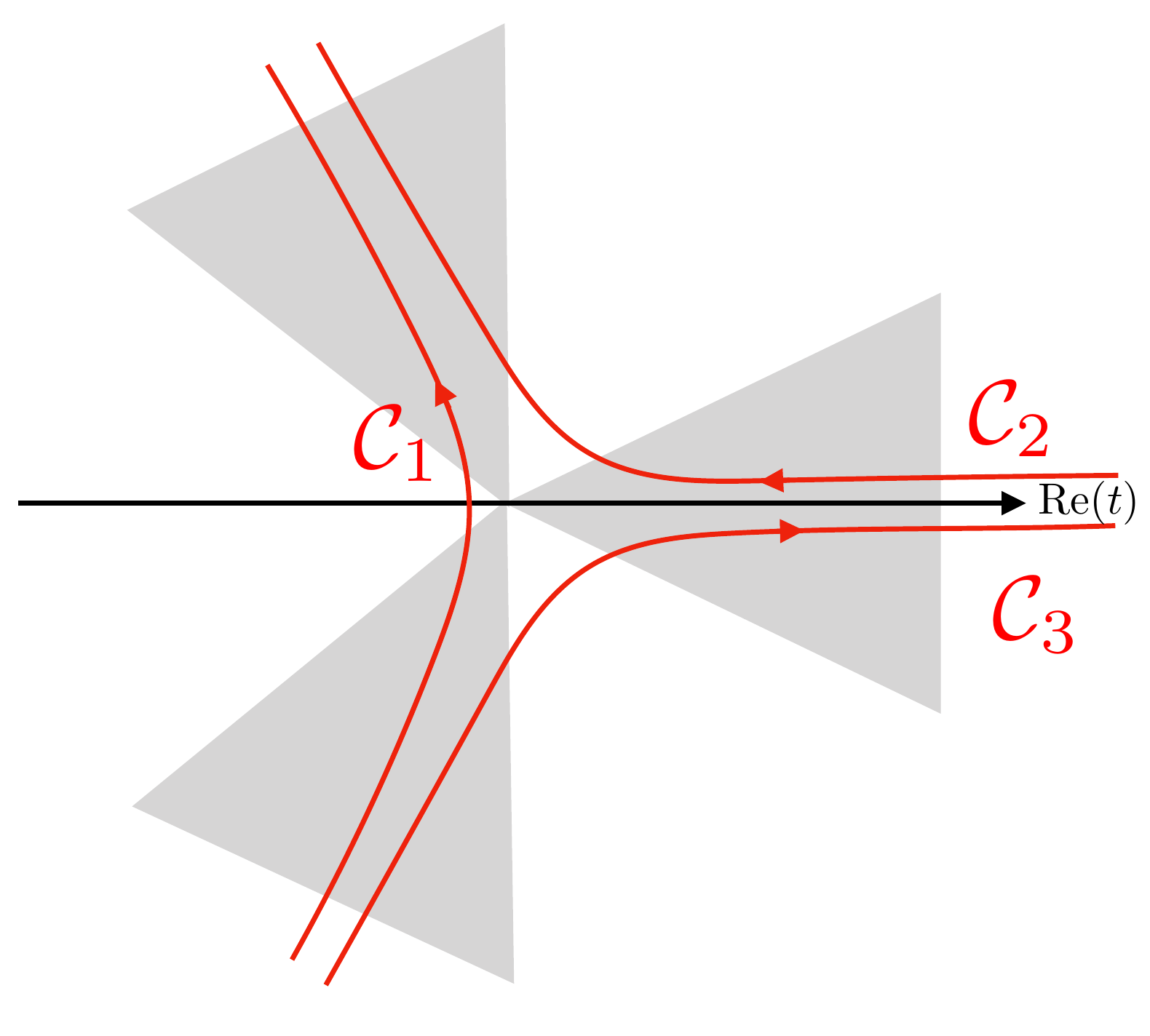}\\
\end{center}
\caption{Choices of contours for the Airy integral. The allowed regions shown in grey with ranges $-\frac{\pi}{6} < \theta < \frac{\pi}{6} $, $\frac{\pi}{2}< \theta < \frac{5\pi}{6}$ and $-\frac{5\pi}{6} < \theta < -\frac{\pi}{2}$. Here we have defined $\theta = {\rm arg}t$. A `good' contour approaches the point at infinity along one of the allowed directions and connects two good regions.}
 \label{fig:AiryContours}
\end{figure}
In this appendix we review a few well-known facts regarding the Airy function and its complex integral representation which we made use of in the main text. Let us define the following integral
\be\label{eq.AiryIntegral}
A_n(z) = \frac{1}{2\pi i}\int_{{\cal C}_n} e^{-\frac{t^3}{3} + z t}dt
\ee
for any of the three contours shown in Figure \ref{fig:AiryContours}. Each of these contours connects one allowed region (shown in grey in Figure \ref{fig:AiryContours}) to another, where these regions are chosen such that the integrand decreases exponentially whenever one approaches infinity inside an allowed region. These are related to the standard Airy functions by
\bea
{\rm Ai}(z) &=& A_1\,,\\
{\rm Bi}(z) &=& i( A_2 - A_3)\,.
\eea
The choice contours and therefore which linear combinations or Airy functions are selected are dictated by the observable we wish to compute. We illustrate this on the example of the integral \eqref{eq.AiryIntegralforDOS} in Section \ref{sec.CausalSymmetryBreakingBulk}, that is
\be
{\cal I} = \int \frac{d\lambda_1 d \lambda_2}{2\pi} e^{-e^{S_0} {\rm STr}  \left(\frac{\lambda^3}{3} + \hat\zeta \lambda \right)}\,,
\ee
where we have the diagonal matrices $\lambda = {\rm diag}(\lambda_1,i\lambda_2)\,, \hat \zeta={\rm diag}(\zeta_1,\zeta_2)$, and the integrations are initially over the real axis. The $\lambda_2$ integral is therefore deformable into the integral along the contour ${\cal C}_1$ and gives the standard Airy function Ai. The second integral, for $\lambda_1$ along the real axis must be deformed into either the contour ${\cal C}_2$ or the contour ${\cal C}_3$. Which contour is chosen depends on the sign of the imaginary part of $\zeta_1$. The correct contour choice is ${\cal C}_3$ for $\zeta_1 + i0$ and ${\cal C}_2$ for $\zeta_1 - i0$. All in all we then find that
\be
{\cal I}_\pm = 2\pi{\rm Ai}(x_2) \left( {\rm Bi}(x_1)\pm i {\rm Ai}(x_1) \right)\,,\qquad x_i := -e^{2S_0/3}\zeta_i\,,
\ee
where the subscript $\pm$ is correlated with the sign of the imaginary part of $\zeta_1$ in a hopefully obvious notation. Then the rest of the analysis in Section \ref{sec.CausalSymmetryBreakingBulk} goes through as advertised. We note that the fact that the contour of integration is chosen according to the imaginary part of $\zeta_i$ is a manifestation of the same mechanism which led to the choice of different saddles in the standard sigma model, but now at the level of the exact Airy integral, which is a special case of the (graded) Kontsevich model studied in Section \ref{sec.CausalSymmetryBreakingBulk}.

We can furthermore use the integral representation \eqref{eq.AiryIntegral} to give a simple description of the asymptotic behavior of the Airy function(s) for large positive and negative values of their argument. In order to obtain the asymptotic behavior of any of the three integrals defined in \eqref{eq.AiryIntegral} one should perform the integral by steepest descent. For this we want to consider the solutions of the saddle-point equation $t^2 - z =0$. For $z>0$ we therefore have two saddle points on the real axis at $\pm \sqrt{z}$, while $z<0$ leads to two saddles along the imaginary axis at $\pm i \sqrt{|z|}$. Which of the saddles contribute to a steepest descent analysis depends on the choice of original contour as well as the sign of the argument. Consider, for example the function Ai$(z)$, that is the contour ${\cal C}_1$. For positive argument the steepest descent contour only passes through the saddle at $-\sqrt{z}$, while the saddle at $\sqrt{z}$ does not contribute. This leads to the well known asymptotics
\be
{\rm Ai}(z) \sim \frac{e^{-\frac{2}{3}z^{3/2}}}{z^{1/4}} \qquad \qquad (z>0)\,.
\ee
However, if $z<0$, the contour ${\cal C}_1$ is deformable into the steepest descent contour through both saddles at $t = \pm\sqrt{|z|}$, which leads to the oscillating behavior
\be
{\rm Ai}(z) \sim \frac{1}{z^{1/4}}  \cos\left(-\frac{\pi}{4} +\frac{2}{3}|z|^{3/2}  \right)  \qquad \qquad (z<0)\,.
\ee
As we commented in the main text the same analysis goes through if we perform a semiclassical expansion of the Kontsevich model for $e^{S_0}\gg 0$, which we can understand  by rescaling the Airy integrand \eqref{eq.AiryIntegral}
\be
e^{-e^{S_0} \left(\frac{\lambda^3}{3} - \zeta \lambda \right) }= e^{-\left(\frac{t^3}{3}- z t\right)}\,,\quad \Leftrightarrow \quad e^{2S_0/3}\zeta = z\,,\,\, e^{S_0/3}\lambda = t\,,
\ee
so that the large $z$ asymptotic analysis just described exactly coincides with the $e^{S_0}\gg 1$ semi-classical evaluation of the integral, as advertised.

\bibliographystyle{utphys}
\bibliography{extendedrefs}

\end{spacing}
\end{document}